\documentclass[11pt,a4paper]{article}
\usepackage[margin=1in]{geometry}   

\usepackage[T1]{fontenc}
\usepackage[utf8]{inputenc}
\usepackage{lmodern}
\usepackage{microtype}

\usepackage{amsmath,amssymb,amsfonts,mathtools}
\usepackage{bm}
\usepackage{graphicx}
\usepackage{booktabs}
\usepackage{subfig}
\usepackage{multirow}
\usepackage{array}
\usepackage{caption}
\usepackage{enumitem}
\setlist{nosep}

\usepackage{xcolor}
\usepackage[
    colorlinks=true,
    linkcolor=black,
    citecolor=black,
    urlcolor=blue
]{hyperref}

\usepackage[numbers,sort&compress]{natbib}

\usepackage{listings}
\usepackage{titlesec}
\titleformat{\section}
  {\large\bfseries}
  {\thesection.}{0.5em}{}
\titleformat{\subsection}
  {\normalsize\bfseries}
  {\thesubsection.}{0.5em}{}
\titleformat{\subsubsection}
  {\normalsize\bfseries}
  {\thesubsubsection.}{0.5em}{}

\titlespacing*{\section}{0pt}{2.0ex plus 0.5ex minus 0.2ex}{1.0ex}
\titlespacing*{\subsection}{0pt}{1.5ex plus 0.4ex minus 0.2ex}{0.7ex}
\titlespacing*{\subsubsection}{0pt}{1.2ex plus 0.3ex minus 0.2ex}{0.5ex}

\makeatletter
\renewcommand{\maketitle}{
    \begin{center}
        {\LARGE\bfseries \@title \par}
        \vspace{1.5em}
        {\large \@author \par}
        \vspace{0.5em}
        {\small \@date \par}
        \vspace{0.5em}
        {\small \itshape Preprint \par}
    \end{center}
    \vspace{1em}
}
\makeatother

\title{Multi-Modal Surface Wave Inversion using Physics Guided Variational Autoencoders}

\author{
Sujith Swaminadhan$^{1}$ \quad
Andrey Bakulin$^{1}$ \quad
Sergey Fomel$^{1}$ \quad
Yangkang Chen$^{1}$\\[0.4em]
$^{1}$Bureau of Economic Geology, \\
Jackson School of Geosciences \\
The University of Texas at Austin \\
Austin, TX 78758 \\
\texttt{sujith.swaminadhan@utexas.edu}
}

\date{October 1, 2026}

\begin{document}

\maketitle

\begin{abstract}
The growth in seismic acquisition techniques has led to large-scale surveys, making traditional manual dispersion analysis logistically and computationally prohibitive. While deep learning techniques can offer an automated alternative, standard convolutional neural networks (CNNs) often fail to account for the non-uniqueness of the inverse problem and can produce geologically unrealistic, smooth velocity models. To address these challenges, we propose a physics-guided multi-modal variational autoencoder (PG-VAE) for scalable near-surface characterization. Our framework introduces a multi-stream fusion strategy that maps independent latent features from both raw shot gathers and spectral-dispersion images by leveraging the physics of Rayleigh-wave dispersion. We implement a specialized decoder architecture with learned transposed convolutions and residual refinement to enforce a structural bias towards layered earth models. Physics guidance is incorporated via a differentiable surrogate network that serves as a forward-consistency regularizer, penalizing predictions that violate the physics of Rayleigh-wave dispersion. The network is trained on diverse synthetic data generated via elastic modeling, and then validated on unseen synthetic data, and subsequently evaluated on an ultra-high-density field dataset from the Devine test site in Texas. We also employ gradient-weighted class activation mapping (Grad-CAM) to examine the input features contributing to the network predictions. The proposed method could serve as a scalable tool for near-surface characterization and as an initial model-building tool for high-resolution elastic full-waveform inversion.
\end{abstract}

\section{Introduction}

Surface waves dominate most of the energy in land seismic data. They can be used to estimate the subsurface shear-wave velocity ($V_s$) profile. Near-surface $V_s$ estimation is a critical task in environmental, geotechnical, and exploration geophysical applications, providing valuable information for characterizing the near-surface. Surface wave inversion (SWI) techniques provide a non-invasive means of deriving these profiles. The foundation of SWI lies in the inherent dispersive nature of Rayleigh waves as they travel through a heterogeneous medium \citep{foti2014surface}. In vertically heterogeneous media, Rayleigh waves propagate with frequency-dependent phase velocities, and different wavelengths probe different depths. Longer wavelengths penetrate deeper, typically up to one to two wavelengths, and are thus sensitive to elastic properties (dominantly $V_s$) at deeper depths. These dispersion trends, i.e., the relationship between phase velocity and frequency, are used to retrieve the near-surface shear-wave structure.

The conventional workflow is a two-step process involving picking the frequency-dependent dispersion curve (DC) from raw seismic data and subsequently modeling and inverting the curve to obtain the corresponding $V_s$ profile \citep{kennett1976inversion, yamanaka1996application, park1999multichannel, luke2007inversion, foti2009non, xia2009high, kallivokas2013site, foti2014surface}. A critical bottleneck in the classical SWI workflow occurs during the initial DC extraction step, as the accuracy of the final $V_s$ model is critically dependent on the quality and reliability of the extracted DCs \citep{mcmechan1981analysis,stoffa1981direct, park1999multichannel, hu2019novel}.

SWI is intrinsically a complex, highly non-linear, ill-posed inverse problem \citep{socco2004surface}. A significant challenge is the non-uniqueness of the solution, where multiple distinct $V_s$ profiles can produce DCs that match the observed data. When analyzing multiple dispersion modes, the misfit function becomes complex and will have numerous local minima \citep{maraschini2010monte}. Traditional local optimization methods, such as linearized inversion schemes, are highly susceptible to local minima, particularly when an accurate starting model is unavailable. Global approaches like Markov chain Monte Carlo (MCMC), simulated annealing (SA), and genetic algorithms (GA) were developed to mitigate the issue of convergence to local minima by systematically searching the entire model space and estimating the posterior probability density (PPD) of acceptable models \citep{brooks2011handbook, song2015grey, cox2016layering}. Most global algorithms, such as MCMC, require many forward simulations to produce a statistically representative estimate of the PPD, making them computationally prohibitive for large-scale inverse problems \citep{maraschini2010monte}. 

Recent advances in deep learning (DL) offer alternatives to address the severe computational and logistical bottlenecks associated with classical SWI. DL models accelerate the process by shifting the computational cost from runtime iterative optimization to an upfront training phase, thereby enabling fast inference \cite[]{shucai2020}. It can also facilitate a transition from the two-step workflow toward a direct inversion approach for $V_s$ profiles. Initial attempts at DL implementations largely focused on the second stage of the workflow, which maps DCs to $V_s$ profiles. \cite{chen2022deep} implemented a convolutional neural network (CNN) to directly invert the fundamental mode of Rayleigh wave DCs for the near-surface $V_s$, and \cite{luo2022constructing} implemented a fully connected neural network (FCN) for the same purpose. \cite{chen2024surface} recently designed a multi-information fusion neural network, which focuses on incorporating parameters such as $V_p$, density, and sensitivity as inputs besides DCs for inverting for $V_s$. These methods remain strictly dependent on the accuracy of the input DC. If the input curves contain mode misidentifications or errors, the network propagates these inaccuracies into the final velocity models. To mitigate the errors associated with manual curve extraction, recent strategies have aimed to automate the process with DL. \cite{chamorro2024deep} proposed a DL methodology using a residual network (ResNet-18) to map seismic shot gathers directly to multiple dispersion modes (fundamental and first-order higher mode), thereby bypassing the need for manual picking. They also incorporated Monte Carlo dropout (MCD) and distributional parameter estimation (DPE) to predict the mean and variance, thereby quantifying the output uncertainty. While this significantly improves automation and accuracy, it only covers the initial DC extraction step rather than full inversion. The predicted curves must still be inverted using traditional algorithms, retaining the computational cost of the final stage.

\cite{aleardi2021hybrid} implemented a residual neural network (ResNet) to map the spectral dispersion image to $V_s$ and layer thicknesses with uncertainty quantification. While promising, this approach highlights a critical limitation of standard CNNs, which are parameterized in a fixed way. The network was restricted to a fixed number of layers, requiring a massive synthetic training set to capture lateral heterogeneity. Generating the necessary synthetic training samples is computationally expensive and requires significant prior knowledge of the site. Despite these advances, a critical gap remains in the implementation of DL for SWI: most networks trained on synthetic datasets exhibit significant domain shift when applied to field data that includes noise, attenuation, and 3D effects \cite[]{shucai2020}. To address these issues, our study integrates physics and robust uncertainty quantification using a novel physics-guided multi-modal variational autoencoder (PG-VAE). While standard DL models offer speed, they often lack rigorous uncertainty quantification. Techniques such as variational autoencoders (VAEs) \citep{kingma2013auto}, physically structured variational inference (PSVI) \citep{zhao2024physically}, and Stein variational gradient descent (SVGD) \citep{liu2016stein} have recently emerged as scalable alternatives to MCMC for efficiently estimating the posterior probability density (PPD). We build upon this Bayesian framework and introduce novel strategies to ensure solutions are physically plausible. We used VAEs to directly map $V_s$ profiles from raw shot gathers and spectral dispersion images. We emphasize enforcing physics constraints using a differentiable surrogate model (a pre-trained Transformer). Our approach also addresses the limitations of traditional smoothing regularization, which can obscure critical subsurface discontinuities. We implement a specialized decoder with learned transposed convolutions and residual refinement. By minimizing mean absolute error (MAE) and gradient penalties, our architecture is designed to recover complex lithological interfaces rather than smooth approximations. This work presents a unified framework that combines the computational efficiency of DL with the physical constraints of numerical inversion, thereby offering a robust, transparent, and geologically realistic approach to near-surface characterization.

The manuscript is organized as follows. We first detail the theoretical foundation and architecture of the proposed PG-VAE framework, including the multi-modal data fusion strategy, the specialized decoder design, and the integration of the differentiable physics surrogate. We then outline the synthetic data generation workflow and evaluate the network's baseline performance, structural fidelity, and uncertainty quantification on an unseen synthetic dataset. We then demonstrate the practical applicability of the framework by employing it on ultra-high-density field data from the Devine Test Site \citep{bakulin2025engineering}. We utilize an explainable AI tool and saliency maps to interpret the network's learned physical attention mechanisms and to provide a geotechnical interpretation of the inverted 2D model. We then assess the
sensitivity of the proposed method to additive noise using controlled synthetic
experiments at different SNR levels. Finally, we discuss the current limitations of our methodology and directions for future research.

\section{Methodology}

This study proposes a PG-VAE method to estimate near-surface shear-wave velocity ($V_s$) and addresses the scalability and manual-method limitations in conventional surface-wave analysis. This framework departs from standard deterministic CNNs by learning a probabilistic latent-space representation of the subsurface, enabling both velocity estimation and uncertainty quantification. We generate diverse synthetic labels with minimal prior geological knowledge. The method is first validated on a synthetic model, which was not used for training, and then applied to a high-density 2D field seismic line. The proposed method achieves improved structural reconstruction and
forward consistency relative to other baseline networks.

\subsection{Network architecture}
The proposed network comprises a dual-stream encoder, a variational bottleneck, and a physics-aware decoder. 
The complete layer-by-layer architectural details, including filter counts, kernel sizes, and strides for the multi-modal encoders and the decoder, are provided in Table \ref{tab:network}.
The network input comprises raw seismic shot gathers, appropriately windowed to contain surface waves, and spectral dispersion images. 
The shot-gather input has dimensions $N_t\times N_r = 401\times201$,
the dispersion image has dimensions $N_v\times N_f^{img} = 260\times24$,
and the output velocity profile contains
$N_z=101$ depth samples.
To extract features from these multi-modal inputs, we employ two independent parallel CNN encoders. The first one is a shot encoder, which processes the raw shot gather of size ($N_t  \times N_r$) where $N_t$ is the number of time samples and $N_r$ is the number of traces. We utilize rectangular convolutional kernels in the initial layers. This anisotropic kernel shape is intended to effectively capture event moveout. The second encoder is the dispersion encoder, which processes the spectral dispersion images or frequency-phase velocity panels of size ($N_v \times N_f^{img}$), where $N_f^{img}$ is the number of frequency samples in dispersion images and $N_v$ is the number of phase velocity samples. We use standard square kernels to capture the curvature and continuity of the dispersion modes. Both encoders employ Batch normalization and LeakyReLU activation functions.
Mathematically, the forward propagation through an intermediate encoder layer $l$ can be expressed as:
\begin{equation}
    y^{(l+1)} = \text{LeakyReLU} \left( \text{BN}( {W}_{\text{enc}}^{(l)} \ast y^{(l)} + {b}_{\text{enc}}^{(l)} ) \right),
\end{equation}
where ${y}^{(l)}\in
\mathbb{R}^{C_l\times H_l\times W_l}$ denotes the input feature tensor to
layer $l$, with $C_l$ input, and $C_{l+1}$ output channels and spatial dimensions $H_l\times W_l$.
The trainable convolution kernel is
${W}^{(l)}_{\mathrm{enc}}\in
\mathbb{R}^{C_{l+1}\times C_l\times k_h^{(l)}\times k_w^{(l)}}$,
where $k_h^{(l)}$ and $k_w^{(l)}$ denote the height and width of the
two-dimensional convolution kernel in layer $l$, respectively,
and ${b}^{(l)}_{\mathrm{enc}}\in
\mathbb{R}^{C_{l+1}}$ is the corresponding channel-wise bias vector.
The operator $*$ denotes the two-dimensional cross-correlation operation. The kernel dimensions, strides,
and number of output channels for each encoder layer are listed in
Table~\ref{tab:network}.

\begin{table}[htbp]
\centering
\footnotesize 
\renewcommand{\arraystretch}{0.9} 
\vspace{0.5em}
\begin{tabular}{lccc}
\hline
\textbf{Network Component} & \textbf{Layer} & \textbf{Kernel/Stride} & \textbf{$C_{out}$} \\
\hline
\textbf{Shot Encoder} & Input & - & 1 \\
& Conv2D + BN + LReLU & $(3\times5), (1, 2)$ & 32 \\
& Conv2D + BN + LReLU & $(3\times3), (2, 2)$ & 64 \\
& Conv2D + BN + LReLU & $(3\times3), (2, 2)$ & 128 \\
& Conv2D + BN + LReLU & $(3\times3), (2, 2)$ & 256 \\
& AdaAvgPool2D & - & 256 \\
\hline
\textbf{Dispersion Encoder} & Input & - & 1 \\
& Conv2D + BN + LReLU & $(3\times3), (1, 1)$ & 32 \\
& MaxPool2D & $(2\times2), (2, 2)$ & 32 \\
& Conv2D + BN + LReLU & $(3\times3), (1, 1)$ & 64 \\
& MaxPool2D & $(2\times2), (2, 2)$ & 64 \\
& Conv2D + BN + LReLU & $(3\times3), (1, 1)$ & 128 \\
& AdaAvgPool2D & - & 128 \\
& Linear & - & 256 \\
\hline
\textbf{Latent Space} & Concatenation & - & 512 \\
& Linear ($\mu, \log(\sigma^2)$) & - & 64 \\
\hline
\textbf{Decoder} & Linear & - & 1664 \\
& Reshape to 1D ($L=13$) & - & 128 \\
& Trans. Conv1D + BN + ReLU & $4, 2$ & 64 \\
& Residual Block 1D (2x Conv1D) & $(3, 1) \times 2$ & 64 \\
& Trans. Conv1D + BN + ReLU & $4, 2$ & 32 \\
& Residual Block 1D (2x Conv1D) & $(3, 1) \times 2$ & 32 \\
& Trans. Conv1D + BN + ReLU & $4, 2$ & 16 \\
& Conv1D (Final $V_s$) & $3, 1$ & 1 \\
\hline
\end{tabular}
\caption{Detailed network architecture for the PG-VAE. The parameters are denoted as (Kernel Size, Stride) for convolutional layers, and $C_{out}$ represents the number of output channels or filters.}
\label{tab:network}
\end{table}

Batch normalization (BN) follows \cite{ioffe2015batch}, and the LeakyReLU
activation \citep{maas2013rectifier} is defined elementwise as
\begin{equation}
\operatorname{LeakyReLU}_{\alpha}(x)
=
\begin{cases}
x, & x\geq 0,\\
\alpha x, & x<0,
\end{cases}
\end{equation}
where $\alpha=0.1$ is the negative-slope parameter
used throughout the encoders. 
The shot encoder uses padding $(1,2)$ in the first convolution and
$(1,1)$ in the remaining convolutional layers, whereas all convolutional
layers in the dispersion encoder use padding $(1,1)$. The two
max-pooling layers in the dispersion encoder use a $2\times2$ kernel
with stride $(2,2)$.

A Global Average Pooling (GAP) layer is included at the end of each encoder to act as a compression scheme. For the final convolutional feature map $y^{(L)}$ with spatial dimensions $H \times W$, the GAP operation compresses each channel $c$ into a single scalar value to form the dense 1D feature vectors $\hat{z}_{\text{shot}}$ and $\hat{z}_{\text{disp}}$:
\begin{equation}
    \hat{z}_{c} = \frac{1}{H \times W} \sum_{i=1}^{H} \sum_{j=1}^{W} y_{c, i, j}^{(L)}.
\end{equation}

After the feature extraction, both branches compress their spatial representations using Global Average Pooling (GAP). While the shot encoder yields a 256-dimensional feature vector directly from the pooling stage, the 128-dimensional output of the dispersion encoder's GAP layer is passed through a fully connected linear layer to project it into an equivalent 256-dimensional embedding space. This ensures both modalities contribute symmetrically to the multi-modal fusion. These two vectors are then horizontally concatenated to form a unified 512-dimensional joint feature representation $\hat{z}_{\text{combined}}$:
\begin{equation}
    \hat{z}_{\text{combined}} = \text{Concat}\left( \hat{z}_{\text{shot}}, \, {W}_{\text{proj}} \hat{z}_{\text{disp}} + {b}_{\text{proj}} \right),
\end{equation}
where ${W}_{\text{proj}}$ and ${b}_{\text{proj}}$ denote the weights and biases of the dispersion projection layer, respectively.

Here,
${W}_{\mathrm{proj}}\in\mathbb{R}^{256\times128}$ and
${b}_{\mathrm{proj}}\in\mathbb{R}^{256}$ are the trainable
parameters of the linear projection, giving
$\hat{z}_{\mathrm{combined}}\in\mathbb{R}^{512}$.
The parameters of the 64-dimensional latent Gaussian distribution are then
obtained using two independent linear projections,
\begin{equation}
{\mu}
=
{W}_{\mu}\hat{ z}_{\mathrm{combined}}
+{b}_{\mu},
\qquad
\log\boldsymbol{\sigma}^{2}
=
{W}_{\sigma}\hat{ z}_{\mathrm{combined}}
+{b}_{\sigma},
\end{equation}

where
${W}_{\mu},{W}_{\sigma}\in\mathbb{R}^{64\times512}$
and
${b}_{\mu},{b}_{\sigma}\in\mathbb{R}^{64}$.

These vectors define the approximate latent distribution
$q(\hat{z}\mid\mathbf{x})=
\mathcal{N}(\boldsymbol{\mu},
\mathrm{diag}(\boldsymbol{\sigma}^{2}))$
\citep{kingma2013auto,doersch2016tutorial}.
A latent sample is then obtained using the reparameterization,

\begin{equation}
    \hat{z}
    =
    \boldsymbol{\mu}
    +
    \boldsymbol{\epsilon}
    \odot
    \exp\left(
    \frac{1}{2}\log\boldsymbol{\sigma}^{2}
    \right),
    \qquad
    \boldsymbol{\epsilon}\sim\mathcal{N}(\mathbf{0},\mathbf{I}),
\end{equation}
where $\odot$ denotes element-wise multiplication.
This probabilistic bottleneck regularizes the inversion and allows uncertainty quantification. We use  $\hat{z}$  for latent variables and $z$ for the depth dimension.

Classical decoders based solely on sequential upsampling operations can
introduce artifacts and may have difficulty recovering sharp structural
variations. To improve the representation of localized velocity contrasts,
we use a decoder that combines learned one-dimensional transposed
convolutions with residual refinement blocks \citep{he2016deep}.
The transposed convolutions progressively increase the depth resolution,
whereas the residual blocks refine the feature representation without
changing its spatial dimension.

The decoder receives a latent sample
$\hat{{z}}\in\mathbb{R}^{64}$.
A fully connected layer first projects the latent vector to a
1664-dimensional representation,
\begin{equation}
    {q}
    =
    {W}_{\mathrm{fc}}\hat{{z}}
    +
    {b}_{\mathrm{fc}},
\end{equation}
where
${W}_{\mathrm{fc}}\in\mathbb{R}^{1664\times64}$ and
${b}_{\mathrm{fc}}\in\mathbb{R}^{1664}$ are trainable parameters.
The resulting vector is reshaped to form an initial feature map
\begin{equation}
    {x}^{(0)}
    \in
    \mathbb{R}^{128\times13},
\end{equation}
corresponding to 128 feature channels and an initial depth-axis length of
13 samples.
The depth dimension is progressively increased using three one-dimensional
transposed-convolution layers. For an upsampling stage $l$,
\begin{equation}
    \tilde{{x}}^{(l)}
    =
    \operatorname{ReLU}
    \left[
    \operatorname{BN}
    \left(
    \mathcal{T}^{(l)}
    \left({x}^{(l)}\right)
    \right)
    \right],
\end{equation}

where $\mathcal{T}^{(l)}$ denotes a learned one-dimensional transposed
convolution. All three upsampling layers use a kernel size of 4, stride of
2, and padding of 1. 
The rectified linear unit (ReLU) activation function
\citep{nair2010rectified} is defined as
\begin{equation}
\operatorname{ReLU}(x)
=
\begin{cases}
x, & x>0,\\
0, & x\leq 0.
\end{cases}
\end{equation}
Residual refinement blocks are applied after the first two upsampling
stages. Unlike the transposed-convolution layers, these blocks preserve both
the number of channels and the depth-axis length. For an input feature map
${x}$ with $C$ channels, the residual branch is

\begin{equation}
    {r}_1
    =
    \operatorname{ReLU}
    \left[
    \operatorname{BN}_1
    \left(
    {W}_1 * {x}
    +
    {b}_1
    \right)
    \right],
\end{equation}

\begin{equation}
    {r}_2
    =
    \operatorname{BN}_2
    \left(
    {W}_2 * {r}_1
    +
    {b}_2
    \right),
\end{equation}

followed by the identity skip connection

\begin{equation}
    {x}_{\mathrm{out}}
    =
    \operatorname{ReLU}
    \left(
    {r}_2+{x}
    \right).
\end{equation}

Here $*$ denotes the one-dimensional cross-correlation operation,
${W}_1,{W}_2\in{R}^{C\times C\times3}$
are trainable convolution kernels, and
${b}_1,{b}_2\in\mathbb{R}^{C}$ are the corresponding
bias vectors. Both convolutions use kernel size 3, stride 1, and padding 1,
so the feature length remains unchanged. 

After the final transposed-convolution stage, the feature tensor has
dimensions $16\times104$. A final one-dimensional convolution with
16 input channels, one output channel, kernel size 3, stride 1, and padding
1 maps these features to a single velocity profile.
The resulting profile contains 104 samples. Because the target velocity
model contains $N_z=101$ depth samples, the output is truncated to the first
101 samples.

\begin{figure}[htbp]
    \centering
    
    \subfloat[Runtime Comparison\label{fig:runtime}]{
        \includegraphics[width=0.51\textwidth]{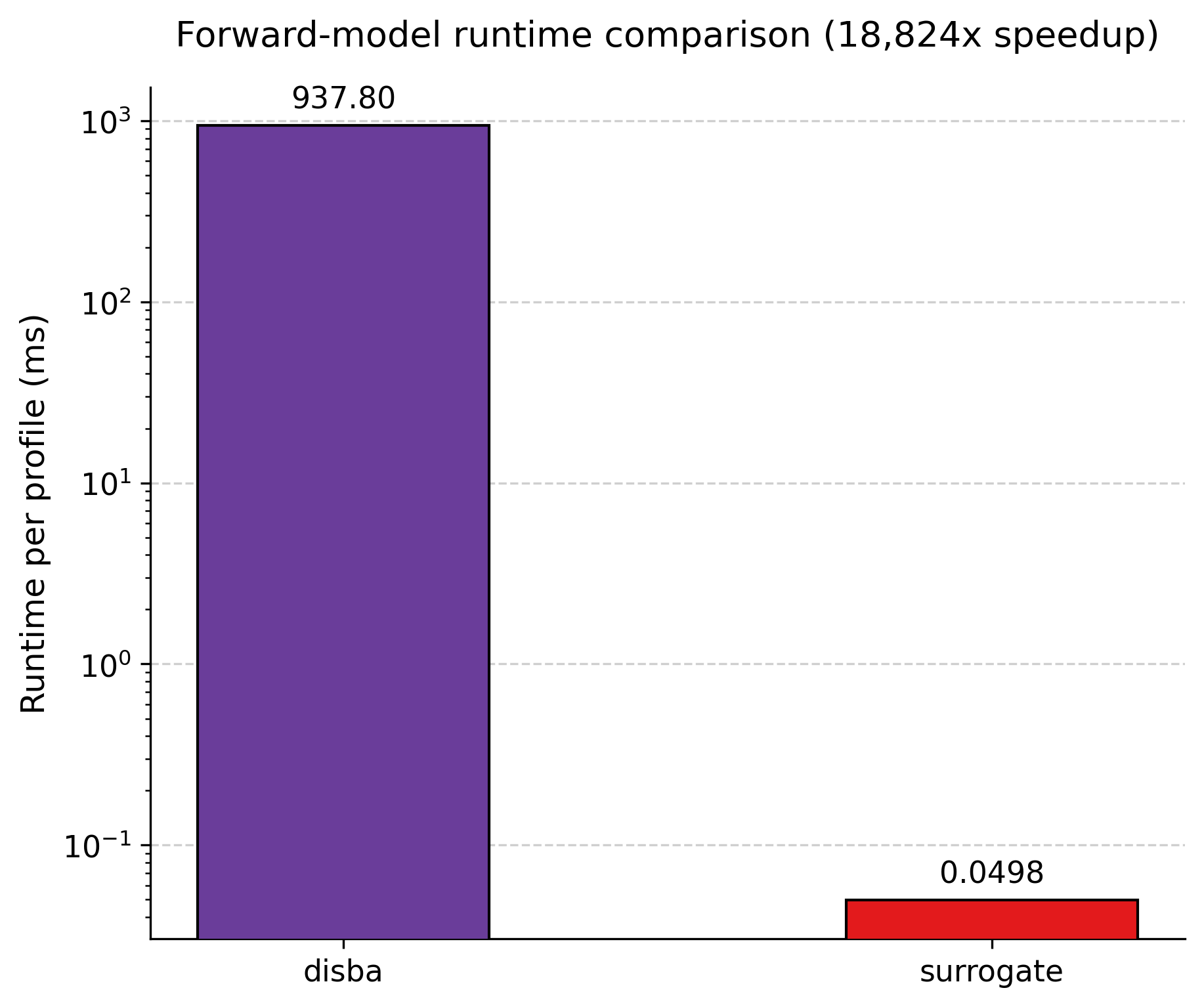}
    }
    \hfill
    \subfloat[Accuracy\label{fig:surrogate-accuracy}]{
        \includegraphics[width=0.43\textwidth]{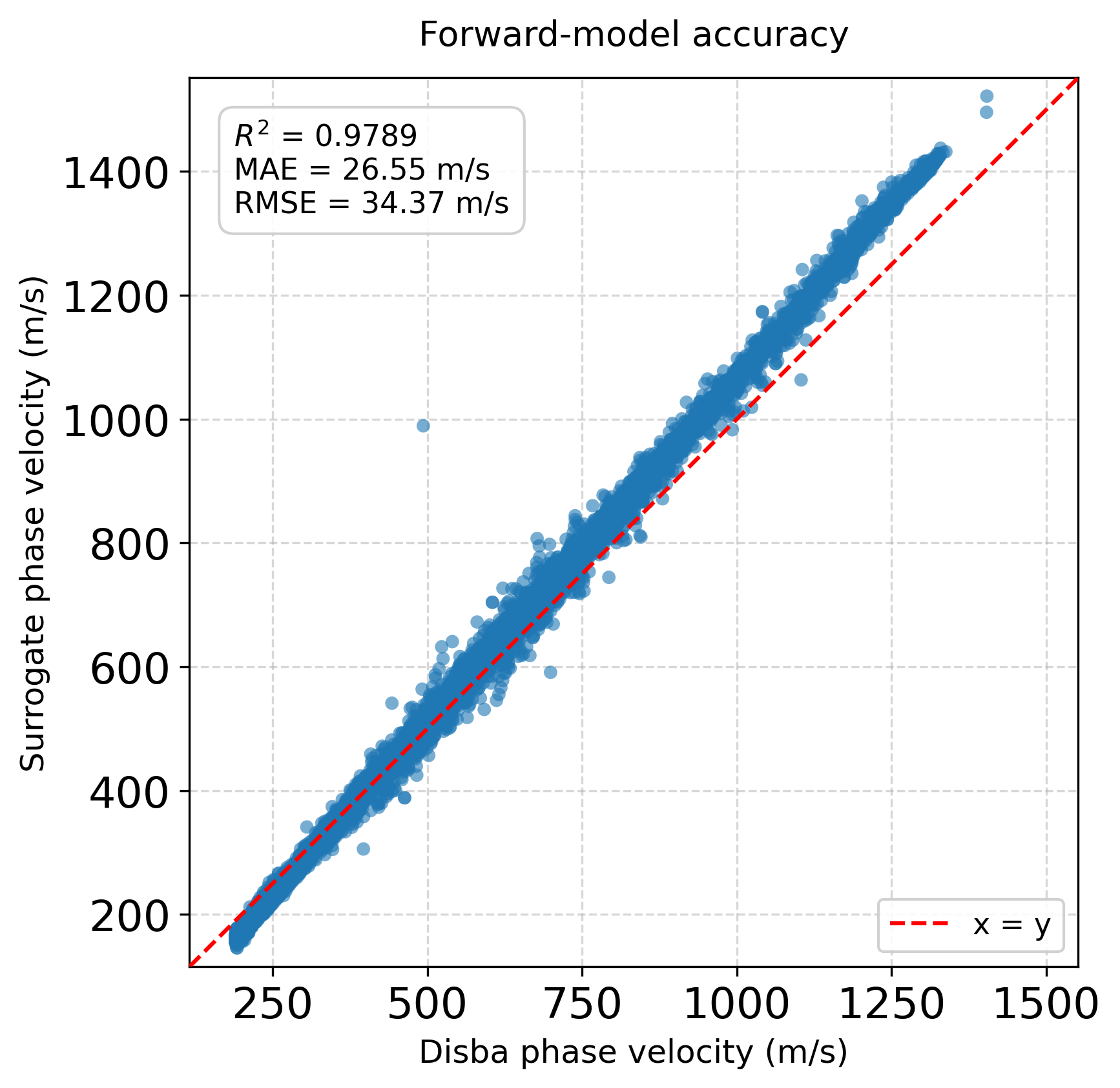}
    }
    
    \caption{Performance evaluation of the differentiable surrogate network. (a) Average computational runtime required to predict the fundamental-mode
    Rayleigh-wave dispersion curve for a single 1D shear-wave velocity profile. The surrogate reduces the average forward-modeling runtime by more than four orders of magnitude relative to \texttt{disba}.
    (b) Crossplot of the fundamental-mode phase velocities predicted by the surrogate for the test set against the corresponding numerical solutions computed using \texttt{disba}. The dashed red line denotes perfect agreement ($y=x$). The surrogate achieves $R^2=0.9789$, an MAE of 26.55~m/s, and an RMSE of 34.37~m/s, demonstrating close agreement with the numerical reference while substantially reducing the forward-modeling cost.}
    \label{fig:runtime,surrogate-accuracy}
\end{figure}

To regularize the generated velocity models toward physical consistency, the decoder output is passed to a differentiable forward operator. We use a pre-trained surrogate Network ($\mathcal{S}$) that approximates the numerical solution of the fundamental Rayleigh wave mode. We train the surrogate network using synthetic labels (shear-wave velocity - phase velocity pairs). The surrogate model is designed to enable learned physics guidance while training the primary VAE network. This requires a differentiable approximation of the forward operator that models the fundamental Rayleigh wave dispersion mode from a given $V_s$ profile. While conventional numerical solvers are accurate, they are mostly non-differentiable and computationally expensive to call iteratively during training. 
To demonstrate both the computational advantage and prediction accuracy of the surrogate network, we compare the average runtime required to compute dispersion curves from $V_s$ profiles using the surrogate and \texttt{disba} \citep{luu_2024_14534395}. 
Figure \ref{fig:runtime,surrogate-accuracy} summarizes the comparison.
In Figure \ref{fig:runtime}, \texttt{disba} requires nearly one second per profile, whereas the surrogate requires only 0.05 ms, a speedup of more than four orders of magnitude. 
The substantial reduction in computational cost is accompanied by
agreement with the numerical reference solution from \texttt{disba}. As shown in
Figure~\ref{fig:surrogate-accuracy}, the surrogate-predicted phase
velocities cluster closely around the $y=x$ line over most of the phase-velocity
range, with slightly increased scatter at the highest phase velocities.
The surrogate achieves an $R^2$ score of 0.9789, a mean
absolute error (MAE) of 26.55~m/s, a root mean square error (RMSE) of
34.37~m/s, and a mean absolute percentage error (MAPE) of 4.35\% relative
to the \texttt{disba} solutions. These results indicate that the surrogate
retains close agreement with the numerical forward model while reducing the
computational cost by more than four orders of magnitude, making it suitable
for repeated evaluation as a differentiable forward-consistency regularizer
during PG-VAE training.

We therefore replaced the numerical forward engine with the neural surrogate $\mathcal{S}$ parametrized by a Transformer
encoder \citep{vaswani2017attention}. $\mathcal{S}$ maps a 1D $V_s$ profile ($v \in R^{N_z}$) to the corresponding fundamental-mode dispersion pick $d_{picks} \in R^{N_f^{pick}}$, where $N_z=101$ is the number of depth samples and $N_f^{pick}=30$ is the number of frequency samples. 
Rather than treating the complete velocity profile as a single flattened
vector, the Transformer processes the profile as a sequence of $N_z$ depth
tokens. Each scalar velocity $v_i$ at depth $z_i$ is first projected to a
128-dimensional embedding using a trainable linear transformation,
\begin{equation}
    e_i = W_{emb} v_i + b_{emb}.
\end{equation}
where $W_{emb}\in\mathbb{R}^{128\times1}$ and $b_{emb}\in\mathbb{R}^{128}$ are the trainable weight matrix and bias vector, respectively. Collecting the embeddings for all depths gives
${E}\in\mathbb{R}^{101\times128}$.
Because self-attention alone is invariant to token ordering, we introduce a
learnable positional embedding
${P}\in\mathbb{R}^{101\times128}$ to retain information about the
depth associated with each velocity sample. The positional embedding is
initialized from a zero-mean Gaussian distribution with standard deviation
0.02 and is optimized jointly with the remaining surrogate parameters. The
input to the first Transformer layer is therefore
\begin{equation}
    h_0 = E + P.
\end{equation}
This element-wise addition helps the model explicitly distinguish between shallow and deep layers during the subsequent attention mechanism.

The core of the surrogate model relies on the scaled dot-product attention mechanism.
The input embeddings $h_0$ are linearly projected into $h$ different subspaces for each head $i \in \{1, \dots, h=4\}$, resulting in query ($Q_i$), key ($K_i$), and value ($V_i$) matrices:
\begin{equation}
    Q_i = h_0 W_{Q_i}, \quad K_i = h_0 W_{K_i}, \quad V_i = h_0 W_{V_i},
\end{equation}
where $Q_h,K_h,V_h\in\mathbb{R}^{N_z\times d_k}$, with $N_z=101$ and $d_k=32$.

The scaled attention-score matrix is
\begin{equation}
S_h =
\frac{Q_hK_h^{T}}{\sqrt{d_k}},
\end{equation}
where $S_h\in\mathbb{R}^{N_z\times N_z}$.
The softmax operation \citep{bridle1990probabilistic} is applied row-wise so that, for each query depth sample $m$, the attention weights over all key depth samples $n$
sum to one:

\begin{equation}
[A_h]_{mn}
=
\frac{\exp([S_h]_{mn})}
{\displaystyle\sum_{j=1}^{N_z}\exp([S_h]_{mj})},
\qquad
m,n=1,\ldots,N_z .
\end{equation}

The output of attention head $h$ is then

\begin{equation}
\operatorname{head}_h
=
A_h V_h
=
\operatorname{softmax}
\left(
\frac{Q_hK_h^{T}}{\sqrt{d_k}}
\right)V_h .
\end{equation}
The outputs from all four heads are concatenated and projected to produce the Multi-Head Self-Attention (MSA) output:
\begin{equation}
    \text{MSA}(Q, K, V) = \text{Concat}(\text{head}_1, \dots, \text{head}_4) W_O,
\end{equation}
where $\text{head}_i = \text{Attention}(Q_i, K_i, V_i)$ and $W_O$ is a learnable output projection matrix. The surrogate consists of two Transformer encoder layers. Each layer uses
an embedding dimension of 128 and four attention heads, giving a
per-head dimension of $d_k=32$.

The integration of Attention-based architectures in geophysics has shown significant promise in capturing long-range dependencies in seismic data \citep{li2024transinver, wang2023seismic, li2025faultvitnet}. The self-attention mechanism enables nonlocal interactions among the
depth-dependent velocity representations, allowing information from different
depth samples to be combined before predicting the complete dispersion curve.
This is useful for surface-wave forward modeling because the phase velocity at
a given frequency generally depends on an extended depth interval rather than
on an individual velocity sample.

After the second Transformer layer, the output
${H}\in\mathbb{R}^{101\times128}$ is flattened to a
12,928-dimensional vector. The projection head consists of a linear
mapping from 12,928 to 128 features, followed by ReLU activation and a
second linear mapping from 128 to the 30 output phase velocities:

\begin{equation}
{g}
=
\mathrm{ReLU}
\left[
{W}_1\mathrm{Flatten}({H})+{b}_1
\right],
\end{equation}

\begin{equation}
\hat{ d}_{\mathrm{picks}}
=
{W}_2{g}+{b}_2,
\end{equation}

where
${W}_1\in\mathbb{R}^{128\times12928}$ and
${W}_2\in\mathbb{R}^{30\times128}$.

\begin{figure}
\begin{center}
    \includegraphics[width=\linewidth]{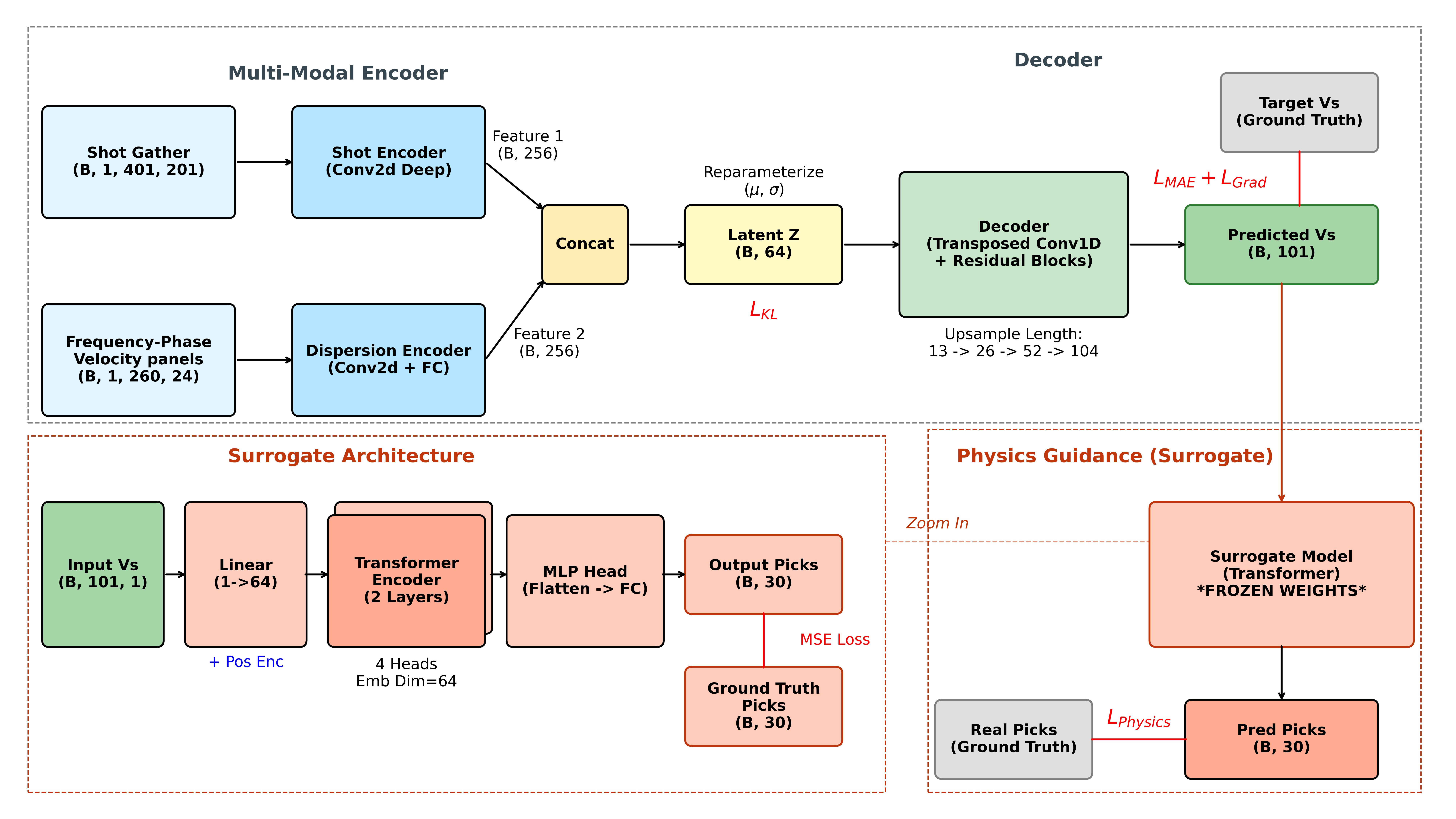} 
    \captionof{figure}{Schematic diagram of the proposed physics-guided multi-modal variational autoencoder (PG-VAE). The network processes raw seismic shot gathers and frequency-phase velocity dispersion panels through parallel 2D convolutional encoders. The extracted feature vectors are concatenated and reparameterized into a probabilistic latent space, which is regularized by Kullback-Leibler divergence ($L_{KL}$). A specialized decoder, utilizing 1D transposed convolutions and residual blocks, upsamples the sampled latent variables to reconstruct the 1D shear-wave velocity ($V_s$) profile. This reconstruction is implemented by minimizing the Mean Absolute Error ($L_{MAE}$) and a structural gradient penalty ($L_{Grad}$). To enforce physics-informed regularization during training, the predicted $V_s$ profile is passed through a pre-trained surrogate model with frozen weights. The surrogate, consisting of a linear projection with positional encoding, a two-layer Transformer Encoder, and an MLP head, maps the velocity profile to predicted dispersion picks. The Mean Squared Error between these predicted picks and the true dispersion picks serves as the physics-guided forward-consistency loss ($L_{Physics}$).}
    \label{fig:network}
\end{center}
\end{figure}

While training the main network, the predicted velocity model  $\hat{v}$ is passed to the pre-trained surrogate network to generate the predicted dispersion picks:
\begin{equation}
    \hat{d}_{picks} = \mathcal{S}(\hat{v}).
\end{equation}
We minimize the error between these predicted picks and the true picks derived from true $V_s$. This loss penalizes predictions that deviate from physically plausible dispersion trends, encouraging the network to generate more geologically realistic models. It is important to note that the surrogate network is used solely as a regularizer during training. During inference, the surrogate model is detached, and only the trained multi-modal encoders and specialized decoder are used to directly predict the $V_s$ profiles from the input data. The summary of the network architecture is given in Figure \ref{fig:network}.

The total loss function $L_{total}$ of our main network is a weighted sum of four components
\begin{align}
L_{\text{total}} &= \lambda_{\text{MAE}} L_{\text{MAE}}
+ \lambda_{\text{grad}} L_{\text{grad}}
+ \beta L_{\text{KL}}
+ \lambda_{\text{physics}} L_{\text{physics}} \label{eq:loss}, \\
L_{\mathrm{MAE}} &= \frac{1}{N_z} \sum_{i=1}^{N_z} |\hat v_i-v_i|, \\
L_{\text{grad}} &= \frac{1}{N_z - 1}
\sum_{i=1}^{N_z - 1}
\left|
(\hat{v}_{i+1} - \hat{v}_i) - (v_{i+1} - v_i)
\right|, \\
L_{\mathrm{KL}} &= -\frac{1}{2N_{\mathrm{lat}}}
\sum_{q=1}^{N_{\mathrm{lat}}}
\left[ 1+\log(\sigma_q^2)-\mu_q^2-\sigma_q^2 \right]. \\
L_{\text{physics}} &= \frac{1}{N_f^{pick}}
\sum_{j=1}^{N_f^{pick}}
\left[S(\hat{v})_j-d_j\right]^2. \label{eq:physics}
\end{align}

Where $L_{MAE}$ is the $L_1$ distance (i.e., mean absolute error) between predicted $V_s$ profile and the true label, and $\lambda_{MAE}$ is the corresponding weight. $L_1$ is preferred over $L_2$ (i.e., mean squared error) to minimize the blurring of edges in the profile \citep{rudin1992nonlinear}. $L_{grad}$ is the structural gradient loss defined by the difference between the difference of the prediction and target, and $\lambda_{grad}$ is its corresponding weight. $L_{grad}$ encourages the correct interface sharpness and depth of the true model. $L_{KL}$ is the KL Divergence \citep{kingma2013auto}, which is the VAE regularization term that enforces the latent representation to approximate a multivariate Gaussian distribution $\mathcal{N}(0,1)$. $\beta$ is the weighting term for KL divergence regularization. $L_{physics}$   is the physics consistency loss and $\lambda_{physics}$ is the corresponding weighting term. In our architecture the value of $\lambda_{MAE}$, $\lambda_{grad}$, and $\lambda_{physics}$ is set to 1 and $\beta$ to gradually increase linearly over epochs from $10^{-5}$ to $10^{-3}$ \citep{burgess2018understanding}. 
In Equation \ref{eq:physics}, the reference variable $d_j$ represents the true dispersion picks. These true picks are computed numerically from the ground-truth 1D $V_s$ profiles using the standard numerical forward-modeling library, \texttt{disba}, prior to training.

With the loss components defined, the final optimization problem seeks to update the weights of the primary VAE network. Let $\phi$ denote the trainable parameters of the multi-modal encoders, and $\theta$ denote the parameters of the physics-aware decoder. The weights of the pre-trained surrogate model, denoted as $\psi$, remain strictly frozen. The optimal network parameters $\phi^*$ and $\theta^*$ are obtained by minimizing the expected total loss over the dataset of $N$ training samples:
\begin{equation}
    (\phi^*, \theta^*) = \arg\min_{\phi, \theta} \frac{1}{N} \sum_{k=1}^N L_{\text{total}}^{(k)}(\phi, \theta; \psi)
\end{equation}

\subsection{Model Interpretability using Grad-CAM}
To evaluate the features prioritized by the multi-modal encoders during inversion, we utilize Gradient-weighted Class Activation Mapping (Grad-CAM) \citep{selvaraju2017grad}. This technique provides a quantitative, spatial assessment of a network's focus by mapping the gradients of a specific output score with respect to the final convolutional feature activations. 
Let $y^c$ denote the scalar output score of interest (the predicted velocity at a specific depth), and let $A^k$ represent the feature map activations of the $k$-th channel in the final convolutional layer of the target encoder. The attention weight $\alpha_k^c$ for that channel is computed via global average pooling of the gradients:
\begin{equation}
    \alpha_k^c = \frac{1}{N_p} \sum_{i} \sum_{j} \frac{\partial y^c}{\partial A_{i,j}^k},
\end{equation}
where $N_p$ is the total number of spatial pixels in the feature map, and $i$ and $j$ denote the spatial dimensions ($N_t$ and $N_r$ for the shot gather or $N_f^{img}$ and $N_v$ for the dispersion image). 

The final coarse heat map of attention, $L_{\text{Grad-CAM}}^c$, is obtained by computing a weighted combination of forward activation maps and passing them through a ReLU operation:
\begin{equation}
    L_{\text{Grad-CAM}}^c = \text{ReLU}\left( \sum_{k} \alpha_k^c A^k \right).
\end{equation}
The ReLU operation retains positive contributions to the target output
score, consistent with the standard Grad-CAM formulation
\citep{selvaraju2017grad}. These attention weights are subsequently mapped back to the input resolution for visual analysis in both synthetic and field experiments.

\subsection{Synthetic labels and training strategy}
\begin{figure}
    \centering
    \includegraphics[width=\linewidth]{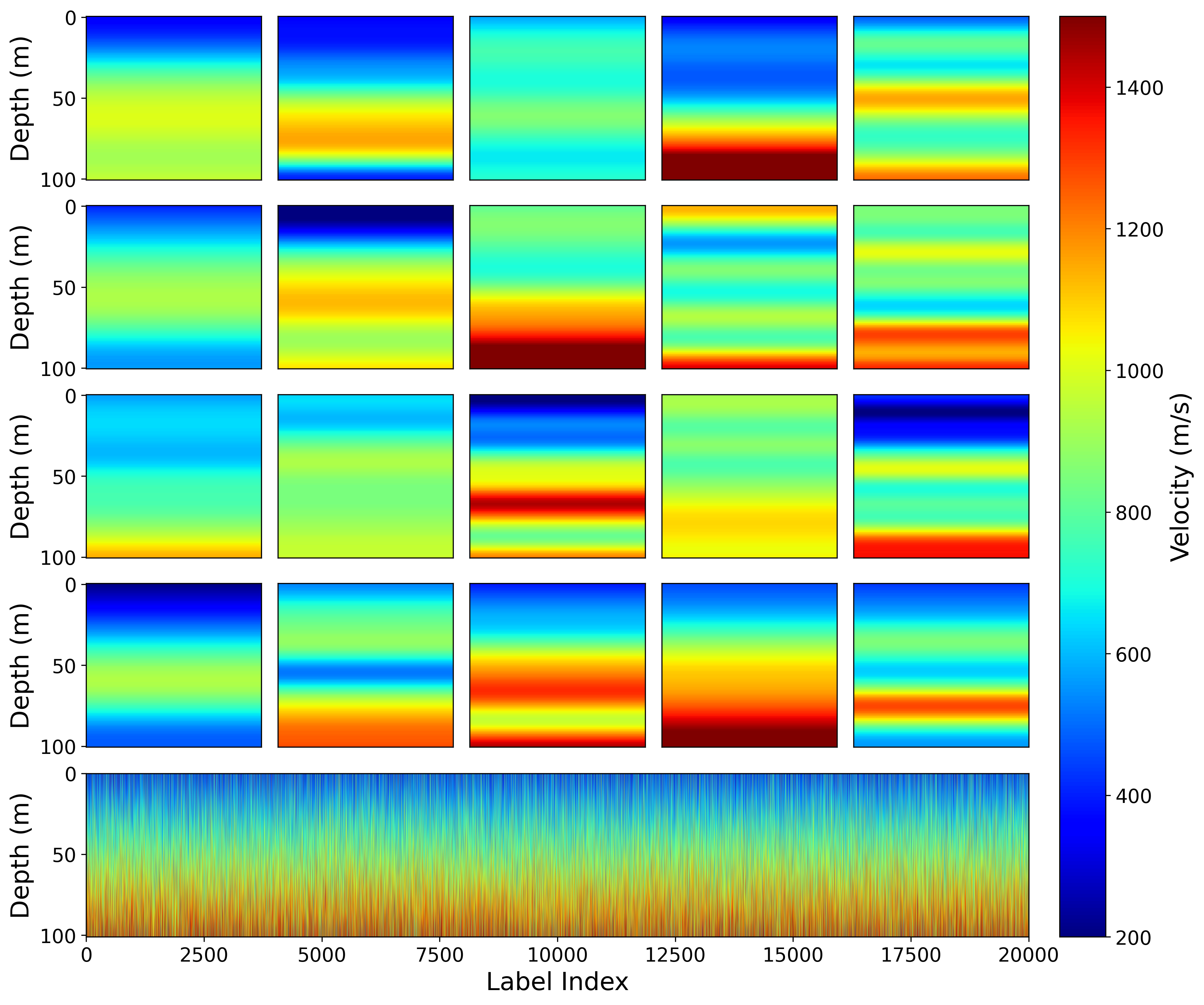}
    \caption{Distribution of 1D shear-wave velocity ($V_s$) profiles utilized for training the surrogate network. The top four rows display 20 randomly selected individual $V_s$ profiles from the dataset. The bottom row shows the complete ensemble of 20,000 synthetic $V_s$ profiles, sorted by label index, illustrating the broad structural variability and velocity ranges encompassed by the training distribution.}
    \label{fig:train-samples}
\end{figure}

To enable physics-guided training, the surrogate network was trained to map depth-dependent shear-wave velocity to frequency-dependent DCs. We generated 20,000 $V_s$ profiles designed to mimic the structural characteristics of near-surface layers. We constructed a 1D linear background gradient that increases with depth and added heterogeneities with strong contrasts. For each profile, the shallow depths are sampled between 200 and 800 $m/s$, and the deepest depths are sampled between 1000 and 1500 $m/s$. The stochastic perturbations with a vertical Gaussian filter mimic physically plausible near-surface heterogeneities and spatial continuity. The amplitudes of the perturbation were randomly scaled to introduce variations of $\pm10 \%$ to $\pm30\%$ relative to the background. The profiles are plotted in Figure \ref{fig:train-samples}, illustrating the structural variability. From these profiles, true DCs are computed by assuming a fixed $V_p/V_s$ ratio of 2 and a bulk density with the empirical relation $\rho = 0.31V_p^{0.25}$ \citep{gardner1974formation} using \texttt{disba}.

\begin{figure}
    \centering
    \includegraphics[width=\linewidth]{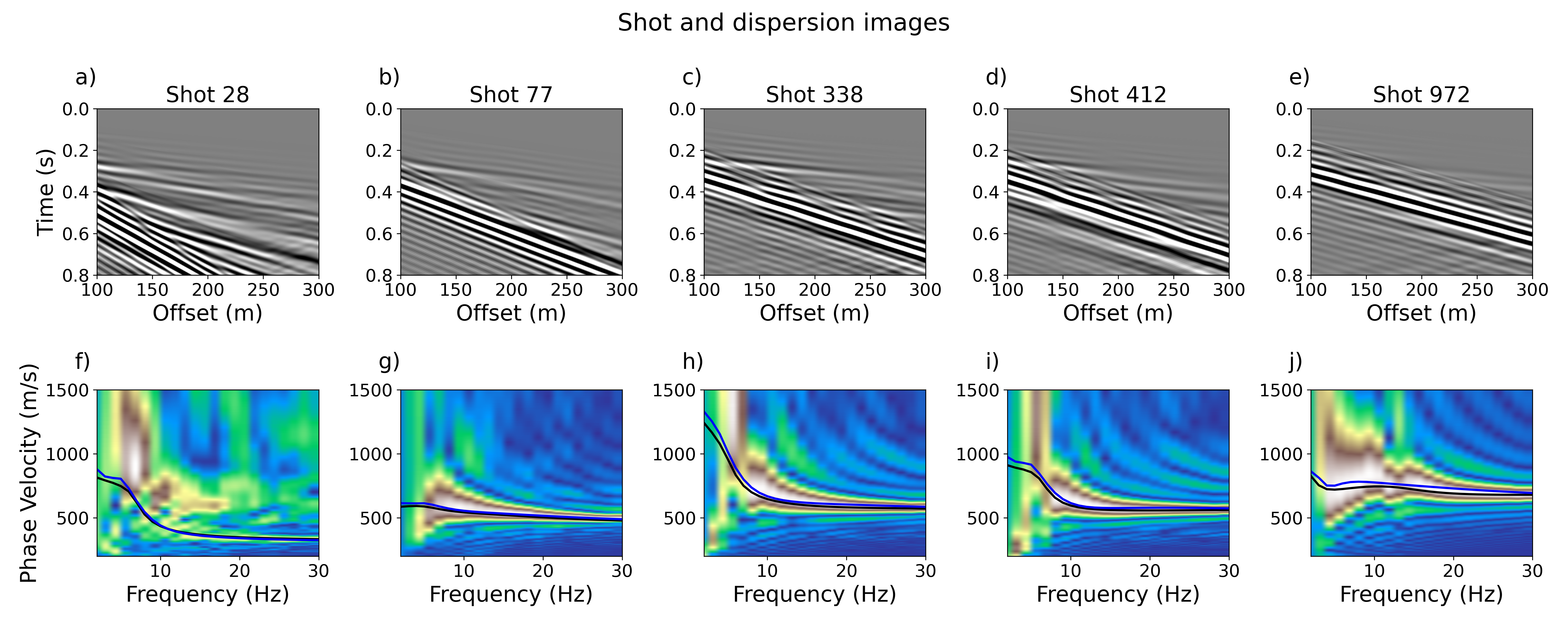}
    \caption{Synthetic training samples generated from diverse $V_s$ profiles. The top row corresponds to the shot gathers from elastic modeling. The modeling was done with an ultra-high-density geometry with 1 m receiver spacing, placed between 100 and 300 m to better capture surface waves. The bottom row shows the spectral dispersion images for each shot, with dispersion curves overlaid. The solid black line represents the conventional numerical solver (\texttt{disba}), and the solid blue line represents the prediction from the trained surrogate network.}
    \label{fig:shot-disp-images}
\end{figure}

To train the main VAE network, we create 1000 $V_s$ profiles using the same strategy. Keeping the same assumptions, we use the empirical relation for P-wave velocity and bulk density. Shot gathers are generated on each 1D layered model through elastic wave propagation. We use devito \citep{devito-api,devito-compiler} to solve the elastic wave equation using the finite-difference method. We limit the offset to 100-300 meters to capture coherent surface-wave events. To mimic a land survey, we use the Klauder wavelet \citep{goupillaud1976signal,brittle2001vibroseis} with a frequency range of 2-30 Hz and a grid spacing of 1 m in both dimensions. The source frequency was band-limited to 30 Hz to mimic field data, in which surface-wave energy is dominant at low frequencies. The receivers are spaced 1 m apart and recorded for 0.8 seconds. Both sources and receivers are placed at the surface. The spectral dispersion image for each shot gather was generated using phase-shift stacking \citep{park1999multichannel}. We consider the frequency range between 5--30 Hz and phase velocity between 200--1500 m/s. The raw shot and corresponding dispersion images constitute the input to the main network, and the true $V_s$ profiles serve as the output labels. The total loss defined in Eq. \ref{eq:loss} was minimized, with the physics loss (Eq. \ref{eq:physics}) defined as the difference between the true DCs and the prediction from the surrogate network. The top row in Figure \ref{fig:shot-disp-images} shows synthetic shot gathers, and the bottom row shows the corresponding spectral dispersion images. These spectral dispersion images are overlaid with the DCs, numerically computed using \texttt{disba} (solid black line) and pre-trained surrogate model prediction (solid blue line). The high degree of alignment among the surrogate predictions, the numerical solutions, and the observed spectral energy trends confirms that the surrogate model achieves sufficient accuracy to serve as a reliable, differentiable physics constraint.

PyTorch \citep{paszke2019pytorch} was used for building the networks. We primarily used the Adam optimizer \citep{kingma2014adam} with an initial learning rate of 0.001 for both the surrogate and main VAE. To ensure stable convergence, the learning rate was dynamically reduced using \texttt{ReduceLROnPlateau} scheduler, with a reduction factor of 0.8 and a patience of 5 epochs, based on the validation loss. The validation set consisted of 20\% randomly selected labels. The surrogate was trained for 200 epochs, and the main VAE for 300 epochs. During training of the main VAE, the surrogate network's weights were strictly frozen, serving solely as a differentiable physics constraint. Both networks were trained with batches of 32 samples. All the inputs are normalized with their mean and standard deviation. For shot gathers and dispersion images, normalization is performed shot by shot and for velocity and DC, a global normalization was implemented.  The training and inference were performed on an NVIDIA RTX A4500 GPU. The surrogate training took approximately 15-16 minutes, and the main network took 8-9 minutes. The inference on 574 shot gathers took approximately 1.5 minutes. Compared with the conventional workflow, which requires time-consuming manual DC picking, this approach can reduce computational time and scale to large-scale surveys. 

Most conventional geophysical inversions rely on a deterministic approach, in which a single solution seeks to explain the geology. However, the ill-posed and non-unique nature of the problem makes the results hard to interpret. By employing a probabilistic framework like a Variational Autoencoder, we map the input seismic data to a continuous multivariate distribution in the latent space rather than to a fixed vector. During inference, we draw $N_s=100$ samples from this learned latent distribution to generate an ensemble of physically plausible $V_s$ profiles. The mean of this ensemble serves as the posterior mean estimate, while the point-wise standard deviation provides a spatially localized measure of uncertainty. This uncertainty quantification is crucial for practical interpretation, as it explicitly highlights deep or highly heterogeneous regions in which the surface-wave data lack sufficient sensitivity to resolve layer boundaries with high confidence.

\section{Results and Discussion}

\subsection{Validation on synthetic 2D model}

We first applied the trained networks to a representative synthetic 2D model to evaluate the performance and accuracy. The model features a 1 m grid spacing in both the X and Z dimensions, with shot and receiver intervals of 1 m. The receivers span offsets from 100 to 300 m from the shot location.

\begin{figure}
    \centering
    \includegraphics[width=\linewidth]{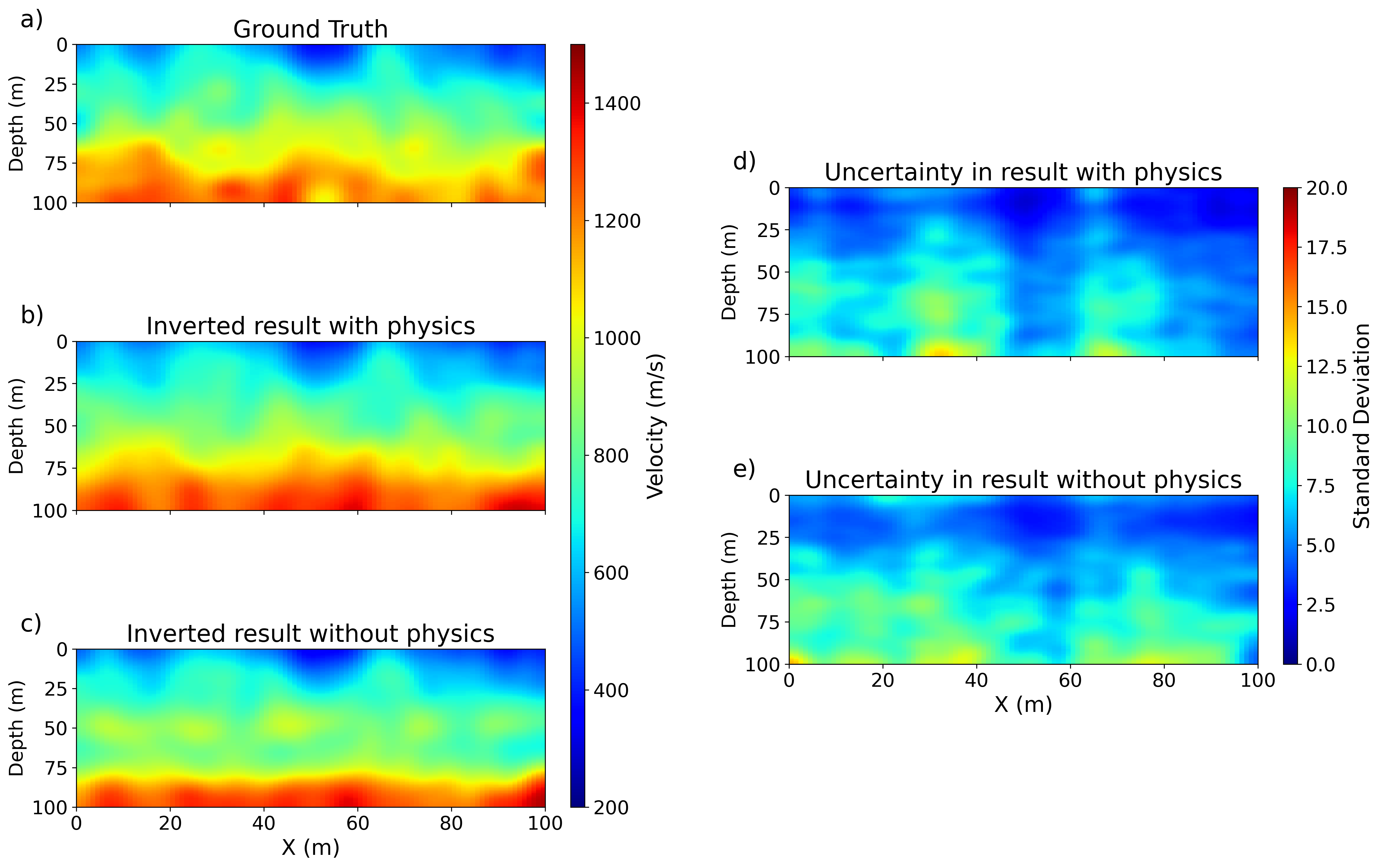}
    \caption{Inversion results on a synthetic 2D model. Panel (a) shows the ground truth $V_s$ profile. Panels (b) and (c) display the mean velocity predictions from the PG-VAE and the baseline VAE, respectively, highlighting the baseline model's failure to capture deep undulations. Panels (d) and (e) show the corresponding standard deviations (uncertainty), demonstrating that the PG-VAE yields higher structural confidence, whereas the baseline VAE exhibits high uncertainty across the flat boundary.}
    \label{fig:Vs-inversion-synth}
\end{figure}

Figure \ref{fig:Vs-inversion-synth} presents a comprehensive comparison between the velocity models derived from the baseline VAE (no physics constraint) and the proposed PG-VAE. Figure \ref{fig:Vs-inversion-synth}a displays the ground-truth model, characterized by a thin, low-velocity shallow layer, an undulating intermediate-velocity zone, and a high-velocity basement at deeper depth. Figure \ref{fig:Vs-inversion-synth}b corresponds to the mean prediction from the PG-VAE, which successfully captures these complex boundaries and preserves the undulations of the deeper layers. In contrast, the baseline VAE (Figure \ref{fig:Vs-inversion-synth}c) exhibits noticeable artifacts and oversimplifies the deeper high-velocity zone into a flat, horizontal layer. While both inversions perform adequately in the shallow subsurface (up to 25 m), the deeper sections highlight a significant contrast in structural fidelity. This divergence is further reflected in the uncertainty maps. Figure \ref{fig:Vs-inversion-synth}d demonstrates that the PG-VAE maintains low uncertainty across most of the profile, with high standard deviations occurring strictly at the bottom left where the mean prediction deviates from the true structure. However, Figure \ref{fig:Vs-inversion-synth}e reveals that the baseline VAE produces high uncertainty across the entire extent of the flat layer at the bottom, indicating higher uncertainty when resolving deep structural geometries.

\subsubsection{Ablation study using single modal inputs}

\begin{figure}
    \centering
    \includegraphics[width=\linewidth]{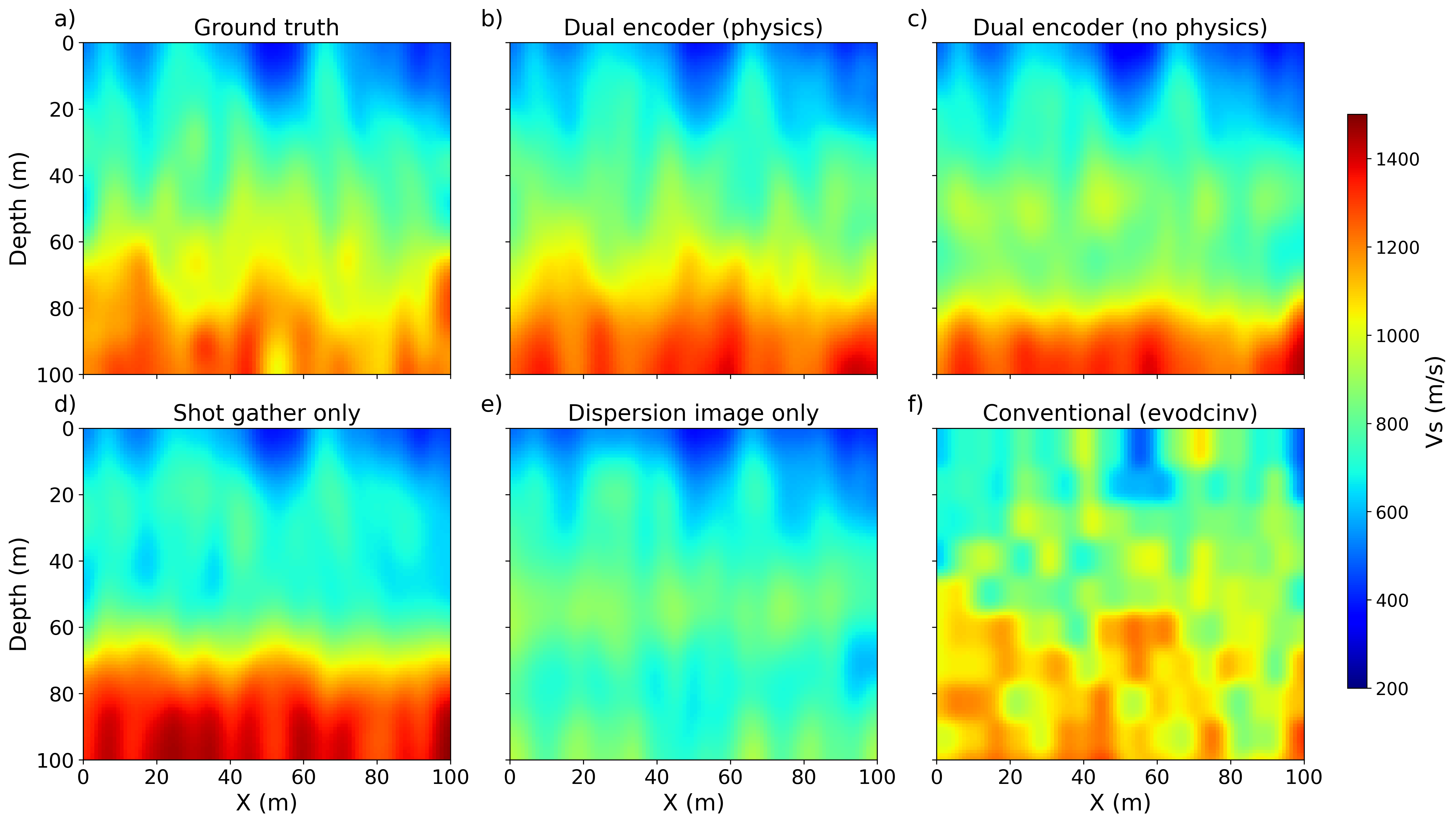}
    \caption{Comparison of mean shear-wave velocity ($V_s$) profiles from the synthetic test set, evaluating the impact of multi-modal inputs and physics guidance. (a) Ground truth velocity model. (b) Inversion result from the proposed physics-guided dual-encoder framework. (c) Baseline dual-encoder result without physics-guided regularization. (d) Single-modality ablation using only raw shot gathers. (e) Single-modality ablation using only dispersion images. (f) Conventional inversion result. }
    \label{fig:ablation}
\end{figure}

To evaluate the necessity and advantage of the multi-modal fusion strategy, we conducted an ablation study utilizing single-modality inputs, restricting the encoder to process either raw shot gathers exclusively or dispersion images exclusively. As observed in the inverted 2D $V_s$ profiles (Figure \ref{fig:ablation}), relying on a single data domain introduces distinct structural limitations and exacerbates the non-uniqueness of the inversion.

The dispersion-image-only model (Figure \ref{fig:ablation}e) yields a heavily smoothed velocity profile that fails to adequately recover deeper, high-velocity structures. Because dispersion images represent an aggregated, macroscopic transformation of the wavefield over a specific spatial window, they inherently lack the high-resolution spatial and phase information preserved in raw waveforms. Consequently, the network captures the background velocity gradient but struggles to resolve sharp geological boundaries.
While raw shot gathers contain rich spatial information, they encompass highly complex, overlapping wavefield phenomena, including body waves, higher-mode surface waves, and scattering. Without the explicit kinematic constraints provided by the dispersion images to isolate the fundamental mode, the shot-gather-only model (Figure \ref{fig:ablation}d) suffers from non-uniqueness. This leads the network to overestimate the shear-wave velocities at deeper zones across the entire profile. 

These ablation results underscore the role of the proposed dual-encoder architecture. By integrating both data streams, the network uses the dispersion images to constrain the global, absolute velocity trend, preventing the overestimation seen in the shot-only model, while leveraging the raw shot gathers to resolve high-frequency spatial boundaries and enforce lateral coherence.

For comparison, the conventional inversion result (Figure \ref{fig:ablation}f) was generated using the \texttt{evodcinv} library \citep{luu_2021_5775743}. The inversion result fails to reconstruct the lateral continuity. While the algorithm maps the broad bounds of the velocity ranges, the resulting 2D profile is degraded by prominent blocky artifacts, structural discontinuities, and spurious vertical striping. This renders the conventional 1D inversion inadequate for high-resolution structural mapping, highlighting the necessity of the laterally coherent constraints learned by the proposed multi-modal network.

\subsubsection{Validation and model interpretability}

\begin{figure}
    \centering
    \includegraphics[width=\linewidth]{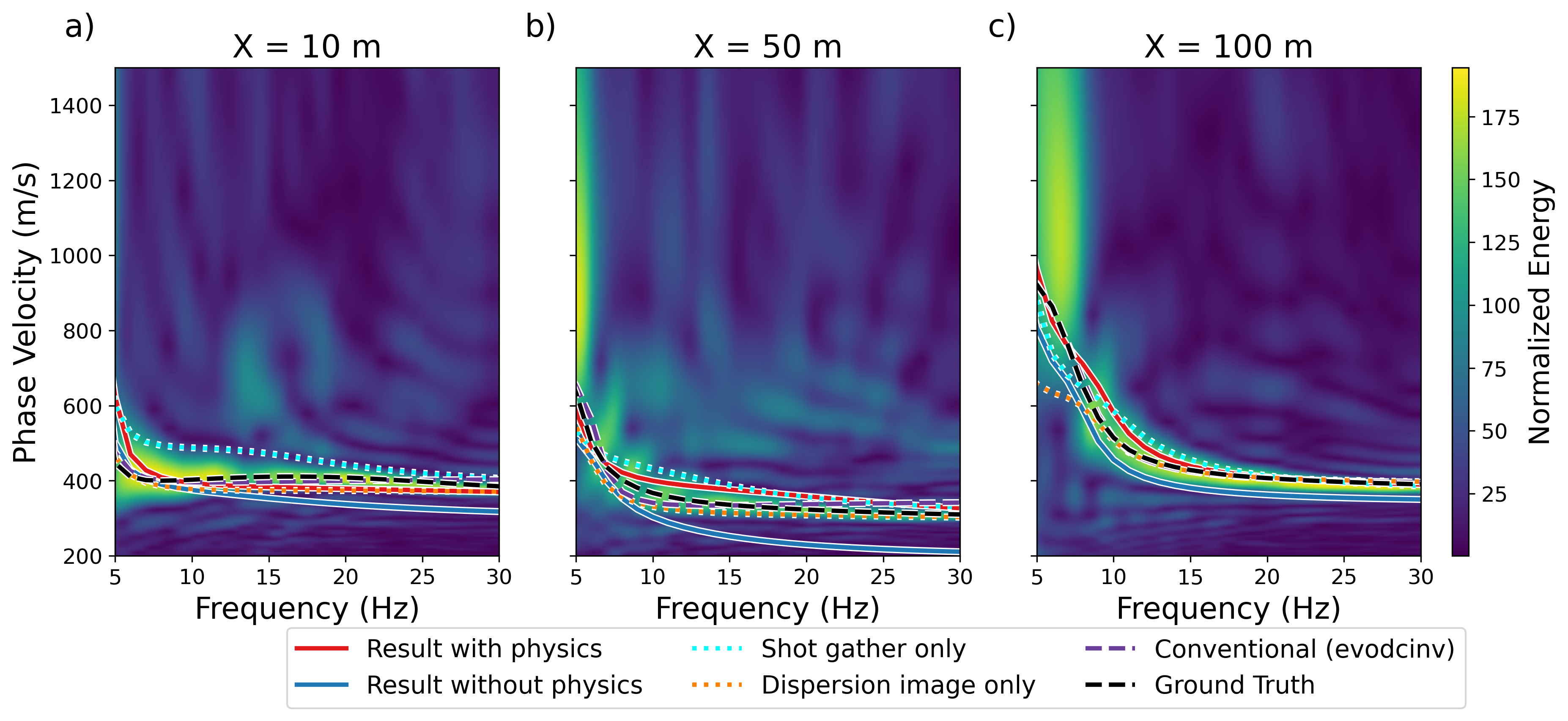}
    \caption{Comparison of predicted and ground-truth fundamental-mode dispersion curves overlaid on the background spectral dispersion images from the synthetic test set. The panels correspond to three representative horizontal locations: (a) X = 10 m, (b) X = 50 m, and (c) X = 100 m. The ground-truth phase velocities (black dashed lines) are compared against predictions from the proposed physics-guided dual encoder (red solid lines), the baseline dual encoder without physics (blue solid lines), the shot-gather-only ablation model (cyan dotted lines), the dispersion-image-only ablation model (orange dotted lines), and the conventional 1D inversion (purple dashed lines). The proposed physics-guided multi-modal approach tracks the ground truth across the entire frequency band, mitigating the velocity overestimation of the shot-only model and the low-frequency underestimation of the dispersion-only model.}
    \label{fig:synth-picks-qc}
\end{figure}

To evaluate the physical validity of the inverted profiles, we forward-modeled the fundamental-mode dispersion curves from the predicted $V_s$ profiles. Figure \ref{fig:synth-picks-qc} compares these modeled curves with the observed spectral dispersion images at three representative spatial locations (X = 10 m, 50 m, and 100 m).
The baseline dual-encoder without physics guidance (Figure \ref{fig:synth-picks-qc}, blue solid lines) deviates from the ground truth (black dashed lines) at frequency extremes. Similarly, the single-modality networks exhibit either systematic velocity overestimation (shot-only, cyan dotted lines) or low-frequency underestimation (dispersion-only, orange dotted lines).
The conventional 1D inversion (purple dashed lines) achieves a tight fit to the ground-truth dispersion curves, particularly at higher frequencies. However, as previously observed in the 2D profiles, this conventional method yields a blocky velocity model. This apparent paradox highlights the severe ill-posed nature and non-uniqueness of surface-wave inversion. The conventional algorithm optimizes each spatial location independently; it converges into disparate local minima. These localized minima yield mathematically equivalent solutions that perfectly reproduce the observed dispersion curve for a single trace but fail to represent the true, laterally continuous geological structure.
The proposed PG-VAE framework (red solid lines) aligns closely with the ground-truth phase velocities across the analyzed frequency band. By integrating a differentiable surrogate network as a forward-consistency regularizer, the PG-VAE constrains the velocity predictions to adhere to surface-wave kinematics. This physics-guided loss mitigates the non-uniqueness observed in the unconstrained baselines, yielding inverted profiles that accurately reproduce the dispersive characteristics of the recorded Rayleigh waves.
While minor visual discrepancies between the physics-guided predictions and the ground truth may still appear, particularly in the low-frequency band (5–10 Hz), these are largely an artifact of the highly nonlinear sensitivity of the Rayleigh-wave forward operator. Because low-frequency surface waves penetrate deeper into the subsurface, even minute variations in the predicted deeper $V_s$ structure translate into magnified changes in the modeled phase velocity, rather than indicating a failure of the inversion.

To quantitatively evaluate the accuracy and structural fidelity of the inverted $V_s$ models, we compute the Mean Absolute Error (MAE), Root Mean Square Error (RMSE), Mean Absolute Percentage Error (MAPE), R$^2$ score, and the Structural Similarity Index Measure (SSIM). 
While the continuous latent space of the PG-VAE recovers the macroscopic lateral coherence of the subsurface, the discrete spatial sampling of the input data introduces minor, high-frequency numerical jitter into the raw predictions. Because the model is characterized by laterally continuous stratified structures, we apply a mild anisotropic Gaussian smoothing filter (2 m standard deviation vertically, 3 m horizontally) to the predicted models shown in Figures \ref{fig:Vs-inversion-synth} and \ref{fig:ablation}. This anisotropic conditioning is applied equally to all models to ensure a fair comparison. Table \ref{tab:metric-synth} presents comprehensive evaluation metrics for both the raw (unconditioned) and smoothed predictions relative to the true synthetic model, demonstrating that the PG-VAE outperforms the baselines in both cases.

\begin{table}[p]
    \centering
    \resizebox{\textwidth}{!}{%
    \begin{tabular}{|c|c|c|c|c|c|}
        \hline
         Model & MAE (m/s) $\downarrow$ & RMSE (m/s) {$\downarrow$} & MAPE (\%) {$\downarrow$} & R$^2$ Score {$\uparrow$} & SSIM {$\uparrow$}\\
         \hline
         PG-VAE (dual encoder, no smoothing) & 118.56 & 170.81 & 14.19 & 0.6872 & 0.6687 \\
         baseline VAE (dual encoder, no smoothing) & 126.8 & 182.98 & 14.51 & 0.6410 & 0.6637 \\
         Shot gather only (no smoothing) & 128.74 & 190.00 & 15.74 & 0.6129 & 0.7004 \\
         Dispersion image only (no smoothing) & 209.53 & 303.34 & 20.73 & 0.0134 & 0.4621 \\
         Conventional (no smoothing) & 255.84 & 329.52 & 32.70 & -0.3099 & 0.3418 \\
         \hline
         PG-VAE (dual encoder, smooth) & \textbf{43.15} & \textbf{60.28} & \textbf{4.63} & \textbf{0.9223} & \textbf{0.7985} \\
         baseline VAE (dual encoder, smooth) & 64.98 & 93.56 & 6.79 & 0.8129 & 0.7478 \\
         Shot gather only (smooth) & 94.61 & 119.00 & 9.88 & 0.6972 & 0.7776 \\
         Dispersion image only (smooth) & 162.98 & 229.91 & 15.59 & -0.1301 & 0.6431 \\
         Conventional (smooth) & 107.11 & 132.35 & 13.82 & 0.6255 & 0.4487 \\
         \hline
    \end{tabular}
    }
    \caption{Quantitative evaluation of the inversion results on the synthetic test set. Metrics are provided for both the raw trace-by-trace network predictions and the final models after applying a 2 m by 3 m anisotropic Gaussian smoothing filter. The PG-VAE dual encoder demonstrates lower error rates and higher structural similarity than the baseline models. The single-modality ablation models confirm the necessity of the multi-modal fusion strategy.}
    \label{tab:metric-synth}
\end{table}

Before any spatial smoothing, the PG-VAE achieves the lowest MAE, RMSE and MAPE and the highest $R^2$ among the compared methods.
The physics-guided network reduces the MAE from 126.80 m/s (baseline VAE) to 118.56 m/s, indicating improved reconstruction accuracy. The necessity of the dual-encoder architecture is evidenced by the ablation models, relying only on shot gathers or dispersion images, which increase the MAE to 128.74 m/s and 209.53 m/s, respectively. The conventional 1D inversion yields the highest error prior to smoothing, exhibiting an MAE of 255.84 m/s and a negative $R^2$ score (-0.3099), which quantitatively reflects the lack of lateral coherence and the presence of artifacts.
The application of the anisotropic Gaussian filter, which suppresses numerical jitter and enforces stratified structures, and improves the quantitative metrics across all results. The smoothed PG-VAE model achieves the lowest overall error, with an MAE of 43.15 m/s (a MAPE of 4.63\%), compared to 64.98 m/s for the smoothed baseline VAE. Structurally, the smoothed PG-VAE attains an $R^2$ score of 0.9223 and a Structural Similarity Index Measure (SSIM) of 0.7985. In contrast, the single-modality networks and the conventional inversion plateau at lower structural accuracies, yielding smoothed $R^2$ scores of 0.6972 (shot-only), -0.1301 (dispersion-only), and 0.6255 (conventional). These comprehensive metrics demonstrate that combining multi-modal inputs with forward-consistent regularization yields a more accurate and structurally faithful velocity model than purely data-driven or traditional methods.

\begin{figure}
\centering
\includegraphics[width=0.8\linewidth]{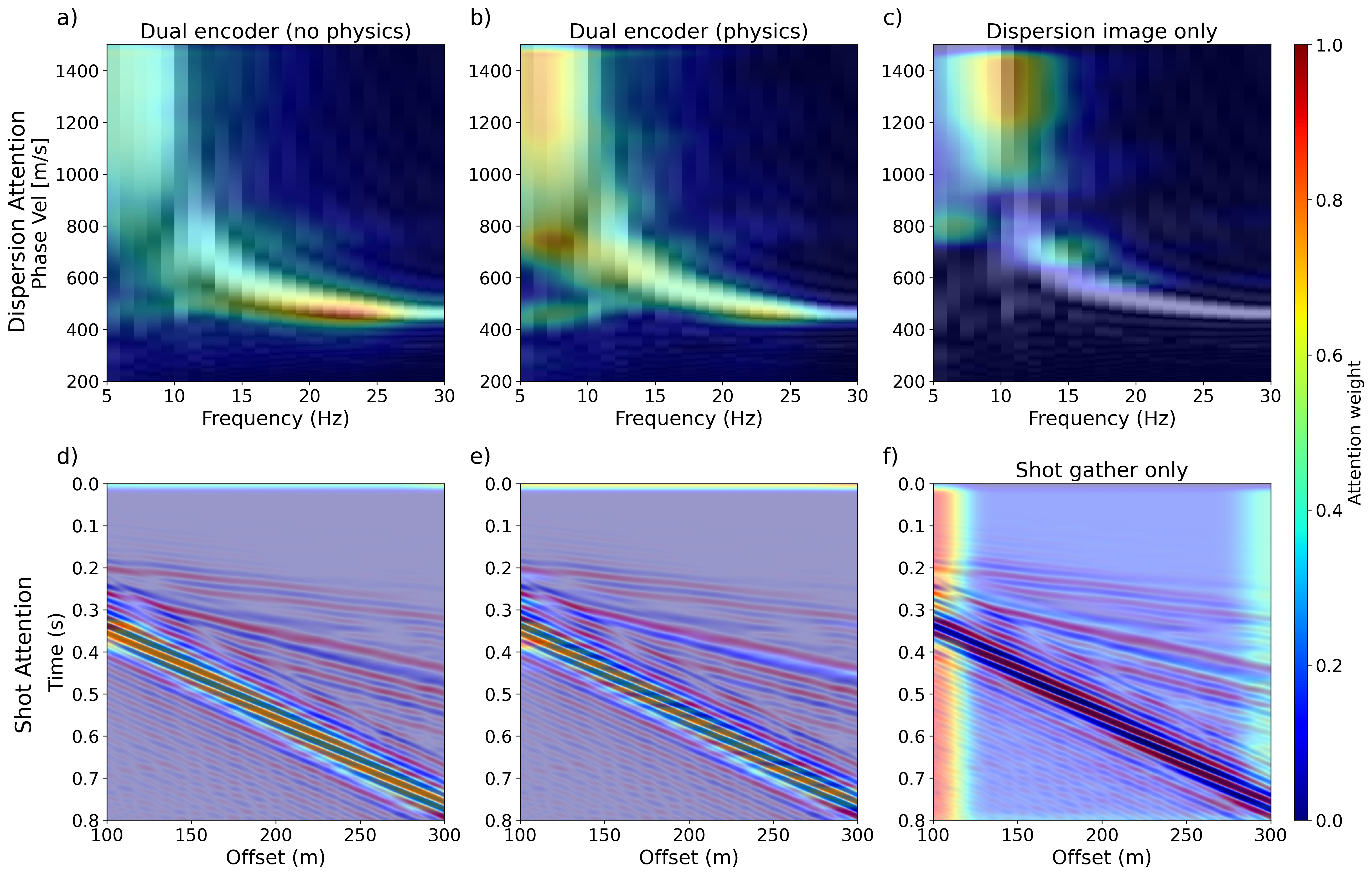}
\caption{Gradient-weighted Class Activation Mapping (Grad-CAM) visualizations of the network attention for a representative example from the synthetic test set. The top row (a--c) displays the attention weights overlaid on the frequency-phase velocity dispersion images, while the bottom row (d--f) shows the attention weights overlaid on the corresponding raw seismic shot gathers. (a, d) Baseline dual encoder without physics-guided regularization. (b, e) Proposed physics-guided dual encoder. (c) Single-modality ablation utilizing only the dispersion image. (f) Single-modality ablation utilizing only the raw shot gather. The physics-guided multi-modal approach (b, e) effectively localizes the network's attention on the fundamental surface-wave mode and the primary surface-wave arrivals. Conversely, the absence of complementary multi-modal constraints results in diffuse, poorly focused attention in the dispersion-only model (c) and introduces anomalous, non-physical vertical artifacts in the shot-only model (f).}
\label{fig:synth-gradcam}
\end{figure}

Figure \ref{fig:synth-gradcam} visualizes the network's attention mechanisms for a representative sample (shot located at X=58 m from the synthetic test set), overlaying the Grad-CAM saliency maps directly onto the input data. In the dispersion domain (top row), the baseline dual encoder trained without physics constraints (Figure \ref{fig:synth-gradcam}a) exhibits a somewhat diffuse attention span around the surface-wave energy. In contrast, the proposed PG-VAE (Figure \ref{fig:synth-gradcam}b) demonstrates a focused and coherent attention mechanism that explicitly tracks the fundamental mode of the dispersion curve. The single-modality dispersion model (Figure \ref{fig:synth-gradcam}c) fails to localize this physical signal, yielding a poorly constrained attention map that smears non-physically across the low frequencies.
In the raw shot gather domain (bottom row), both the baseline dual encoder (Figure \ref{fig:synth-gradcam}d) and the PG-VAE (Figure \ref{fig:synth-gradcam}e) display narrow, focused attention stripes that align with the linear moveout of the surface waves. This suggests that the multimodal architectural design can capture the spatiotemporal coherence of the propagating wavefront. However, when the network is restricted to processing only the shot gathers (Figure \ref{fig:synth-gradcam}f), it develops non-physical vertical attention artifacts that span across all time samples. These visualizations reveal that by the proposed approach compel the encoders to learn the correct set of physical features necessary to accurately invert the subsurface structure.

\subsection{Inference on field data}

\begin{figure}
    \centering
    \includegraphics[width=\linewidth]{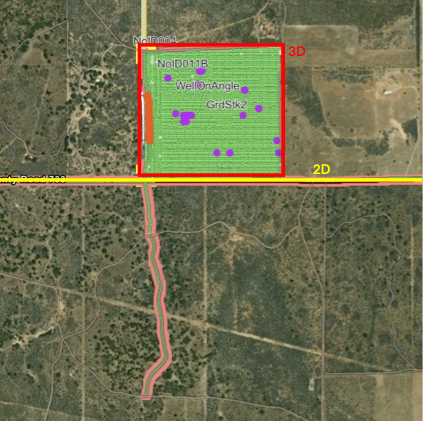}
    \caption{Map of the acquisition geometry at the Devine Test Site, Texas. The yellow line marks the ultra-high-density E-W 2D profile used to validate the performance of the network. The 2D line has over 600 shots at 5 m spacing, and the receivers are spaced 2.5 m apart over 4 km. The red rectangular box corresponds to the adjacent 3D survey. Map modified from \cite{bakulin2025engineering}.}
    \label{fig:map}
\end{figure}

After validating the network on the synthetic model, we deploy the framework to characterize a real 2D survey. The data are from the Devine Characterization Survey, conducted in late 2024 by the Bureau of Economic Geology (BEG) as part of a hydrogen injection test initiative \citep{bakulin2025instrumenting}. A complete description of the survey design and a preliminary data analysis are available in \cite{bakulin2025ultra,bakulin2025engineering}. Figure \ref{fig:map} illustrates the survey geometry, which consists of two long-offset 2D lines extending East-West and North-South, along with a dense 3D survey marked by the red rectangle. For this study, we specifically focus on the East-West 2D line (marked in yellow). This ultra-high-density 2D line comprises 666 shot points and over 1,300 geophones, yielding a maximum offset of 4 km. The acquisition geometry features a dense receiver spacing of 2.5 m and a shot spacing of 5 m. For the inference phase, we utilized 574 shots starting from the western edge of the line.

To align the field data with the input distribution of our trained networks, we applied a bandpass filter matching the synthetic training frequency band of 2 to 30 Hz. The data were windowed to an offset range of 100--300 m and truncated to a maximum recording time of 0.8 s. Because the field data were recorded at a 2.5 m receiver interval, while the networks were trained on a 1 m spacing, we spatially interpolated the real shot gathers to match the required 1 m receiver spacing.
This interpolation also improves the coherency of the surface waves.
These preprocessed shot gathers and their corresponding spectral dispersion images served as the inputs for the inversion.

\begin{figure}
    \centering
    \includegraphics[width=\linewidth]{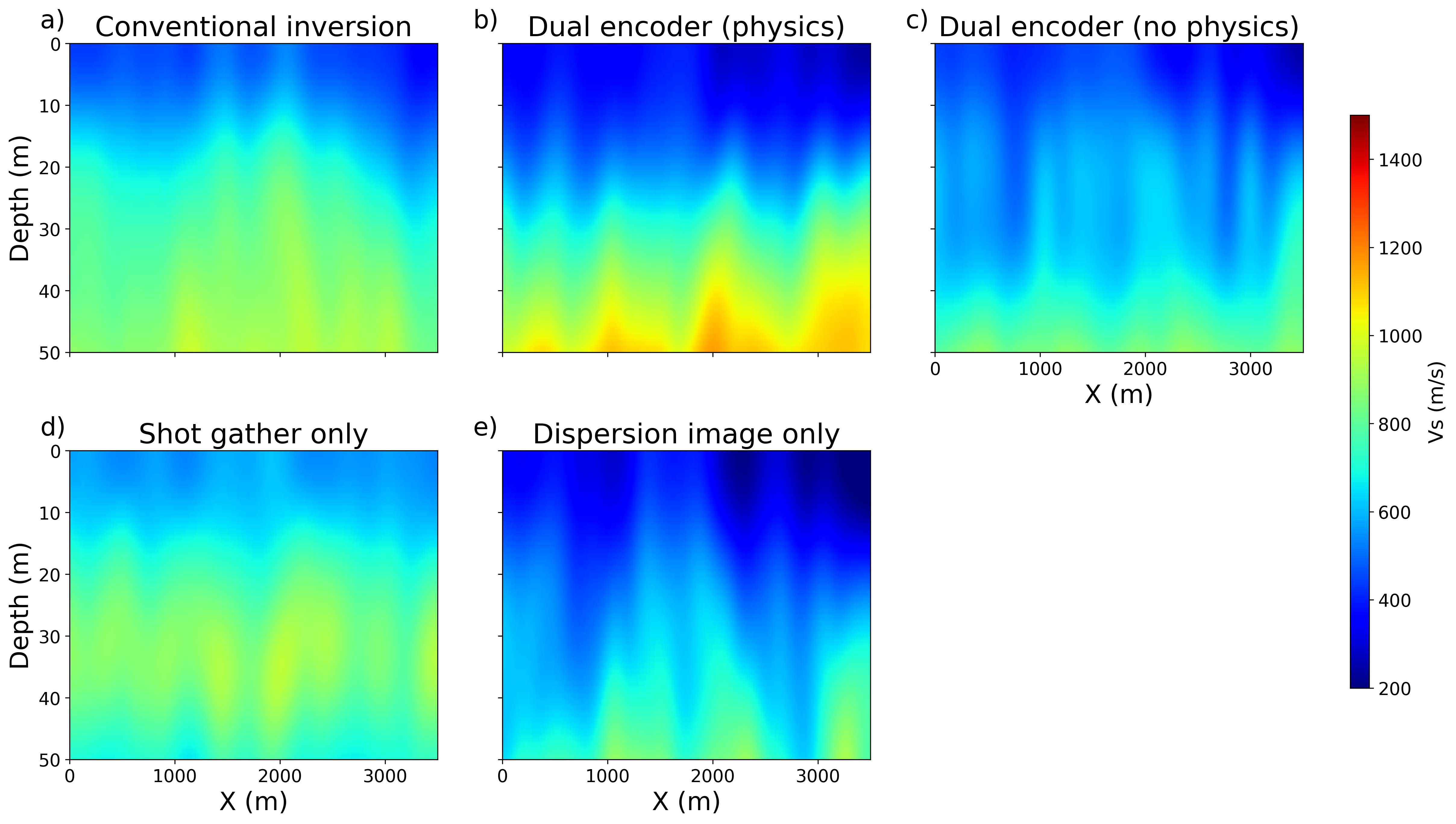}
    \caption{Comparison of inverted mean 2D shear-wave velocity ($V_s$) profiles for the field dataset. (a) Conventional inversion result. (b) Inversion result from the proposed physics-guided dual-encoder (PG-VAE) framework. (c) Baseline dual-encoder result trained without physics-guided regularization. (d) Single-modality ablation utilizing only raw shot gathers. (e) Single-modality ablation utilizing only dispersion images. Consistent with the synthetic validation, the single-modality networks exhibit artifacts on field data. The shot-gather-only model (d) overestimates near-surface velocities, while the dispersion-only model (e) and baseline multi-modal network (c) fail to resolve the high-velocity bedrock at depth. While the conventional inversion (a) recovers the macroscopic gradient, the proposed PG-VAE (b) resolves a structurally sharper and geologically plausible high-velocity bottom interface, demonstrating the necessity of combining multi-modal inputs with physics regularization for real-world applications.}
    \label{fig:ablation-real}
\end{figure}

Figure \ref{fig:ablation-real} compares the inverted 2D shear-wave velocity ($V_s$) 
profiles obtained from the conventional workflow, the proposed PG-VAE, and the baseline 
VAE network, and the single-modality ablation models when applied to the real field 
dataset.
The conventional 1D inversion (Figure \ref{fig:ablation-real}a) recovers the broad 
background velocity trend, but its independent trace-by-trace optimization yields a laterally discontinuous model that lacks coherent representation of deeper structures. 
The proposed PG-VAE (Figure \ref{fig:ablation-real}b) recovers a more detailed and geologically plausible structural profile. By enforcing kinematic consistency through the differentiable surrogate model, the network regularizes the multi-modal latent 
representation toward forward-consistent velocity models, delineating a shallow low-velocity zone that 
transitions across an undulating boundary into intermediate strata, and recovering a 
deeper high-velocity interface exceeding 1200 m/s that the other methods struggle to 
resolve. 
The baseline dual-encoder VAE without physics guidance 
(Figure \ref{fig:ablation-real}c) captures the broad velocity gradient but struggles to resolve sharp structural boundaries. Because the field data contain complex noise, 
scattering, and out-of-plane effects absent from the synthetic training set, the 
unconstrained network generalizes less effectively than the physics-regularized 
counterpart. 
The single-modality networks exhibit distinct calibration and resolution issues under 
field conditions. The shot-gather-only model (Figure \ref{fig:ablation-real}d) tends 
to overestimate near-surface velocities,
whereas the dispersion-only model (Figure \ref{fig:ablation-real}e) produces an oversmoothed profile that does not resolve the deeper high-velocity zone.

\begin{figure}
    \centering
    \includegraphics[width=0.6\linewidth]{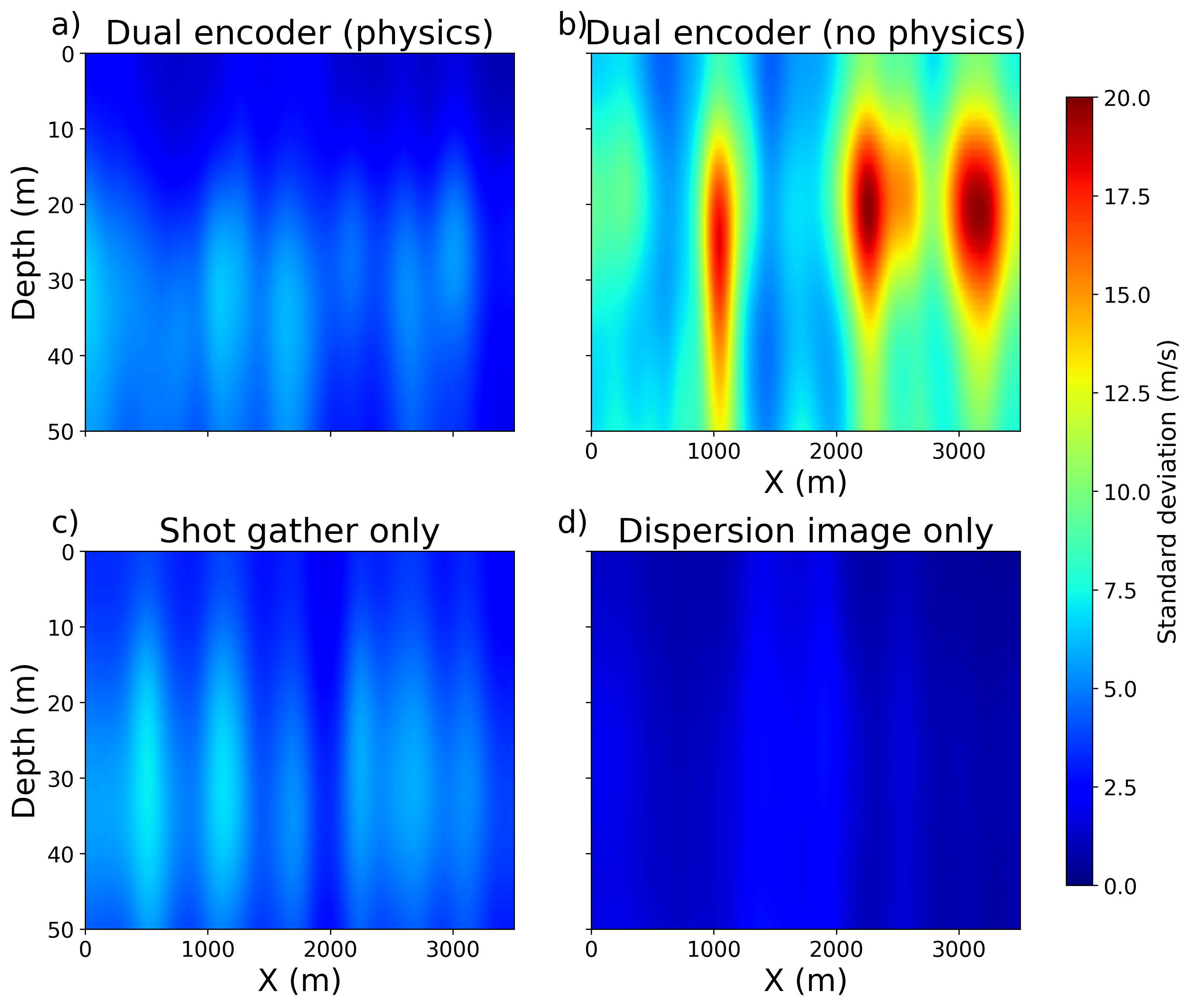}
    \caption{Estimated uncertainty maps of the inverted 2D shear-wave velocity profiles for the field dataset. The maps quantify the variation of the generated $V_s$ models across multiple latent space samples. (a) Proposed physics-guided dual encoder (PG-VAE). (b) Baseline dual encoder trained without physics regularization. Single-modality ablation utilizing (c) only raw shot gathers and (d) only dispersion images. The unconstrained baseline network (b) exhibits anomalies, particularly at greater depths. By enforcing kinematic consistency through the surrogate model, the PG-VAE (a) suppresses these high-variance zones, yielding a stable, confident structural prediction. While the single-modality ablation model (d) displays uniformly low variance, it fails to resolve deep structures. Its low standard deviation is therefore attributable to high model bias and macroscopic over-smoothing rather than to predictive fidelity.}
    \label{fig:uncertainty-real}
\end{figure}

Figure \ref{fig:uncertainty-real} maps the statistical stability of each model by 
computing the point-wise standard deviation of $V_s$ profiles sampled across the 
learned latent distribution. The proposed PG-VAE (Figure \ref{fig:uncertainty-real}a) 
maintains a generally low and spatially stable uncertainty profile. The shallow zones 
are resolved with high statistical confidence, while a relative increase in standard 
deviation is observed in the deeper sections, predominantly on the left flank of the 
profile where surface-wave sensitivity diminishes. The unconstrained baseline dual 
encoder (Figure \ref{fig:uncertainty-real}b) exhibits localized uncertainty anomalies, 
with standard deviation peaking near 20 m/s, reflecting the 
instability that arises in the absence of physical regularization when the network 
encounters field data conditions outside its training distribution. 
The shot-gather-only model (Figure \ref{fig:uncertainty-real}c) displays uniformly 
low uncertainty. However, as established in the structural evaluation above, this 
network produces biased velocity predictions, and its low variance reflects 
consistent reproduction of a biased estimate rather than predictive confidence. 
Similarly, the dispersion-image-only model (Figure \ref{fig:uncertainty-real}d) 
exhibits uniformly low standard deviation. This low variance is likely a symptom of 
underfitting rather than accuracy, as the network predicts the same smooth background 
trend regardless of local data variations and fails to resolve deeper 
structural boundaries.

\begin{figure}
    \centering
    \includegraphics[width=\linewidth]{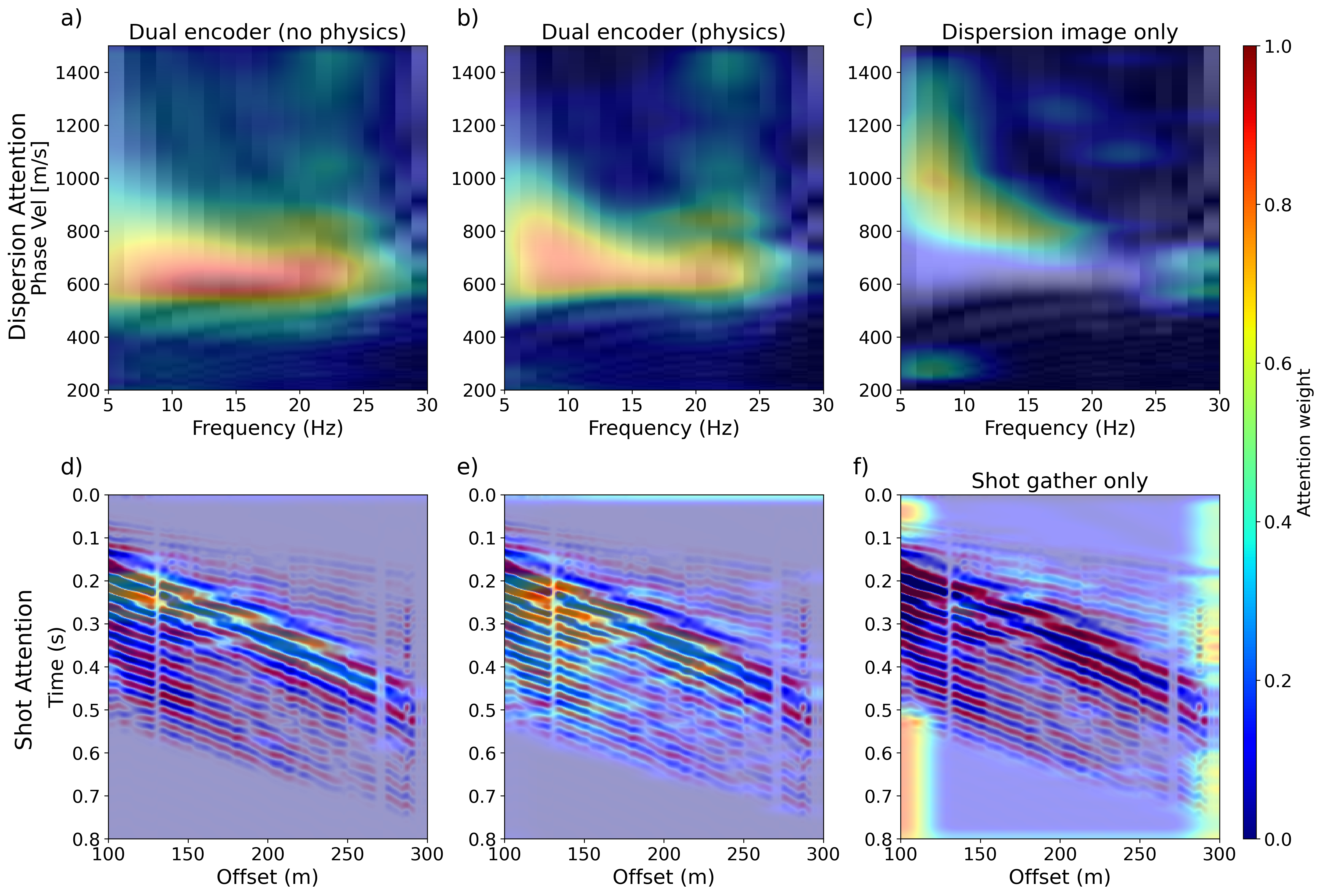}
    \caption{Grad-CAM visualizations of the network attention for a representative sample from the field dataset. Panel layout and baseline descriptions are identical to Figure \ref{fig:synth-gradcam}. The PG-VAE (b, e) helps focus its attention on the fundamental surface-wave mode and the surface wave arrivals. In contrast, the baseline dual encoder shows diminished focus. The dispersion-only model (c) exhibits diffuse attention that fails to track the physical dispersion trend, while the shot-only model (f) becomes smeared by non-physical vertical artifacts at the near and far offsets.}
    \label{fig:real-gradcam}
\end{figure}

To investigate how the networks handle the transition from synthetic training data to complex field observations, we analyze the Grad-CAM saliency maps in Figure \ref{fig:real-gradcam}. Without physics guidance, the baseline dual encoder exhibits diminished focus when exposed to the complexities of field data. In the dispersion domain (Figure \ref{fig:real-gradcam}a), its attention becomes relatively diffuse, and in the shot domain (Figure \ref{fig:real-gradcam}d), it captures the primary arrivals but is influenced by background noise. The unconstrained single-modality networks also demonstrate significant vulnerability when applied to field data. The dispersion-only model (Figure \ref{fig:real-gradcam}c) struggles to localize the coherent dispersion trend, while the shot-only model (Figure \ref{fig:real-gradcam}f) overfits to non-physical vertical artifacts caused by trace irregularities at the array edges. 
Conversely, the PG-VAE demonstrates more focused feature extraction. Its attention in the dispersion domain (Figure \ref{fig:real-gradcam}b) more consistently follows the high-energy fundamental mode, and in the shot domain (Figure \ref{fig:real-gradcam}e), it remains concentrated on the coherent surface-wave moveout despite the presence of background field noise. These visualizations suggest that the forward-consistency loss acts as an effective regularizer.

\begin{figure}
    \centering
    \includegraphics[width=\linewidth]{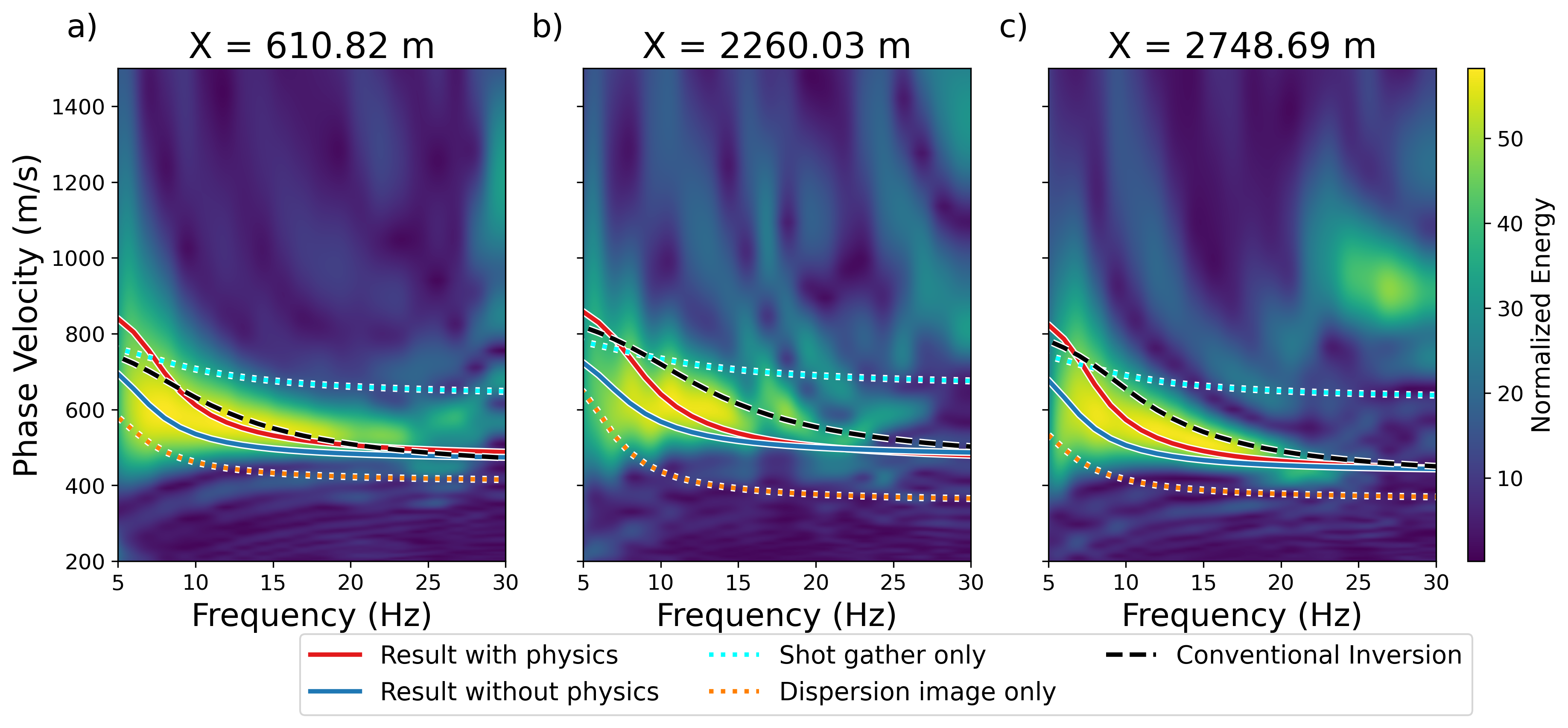}
    \caption{Comparison of predicted fundamental-mode dispersion curves overlaid on the observed spectral dispersion images for the real field dataset. The panels correspond to three representative horizontal locations: (a) $X = 610.82$ m, (b) $X = 2260.03$ m, and (c) $X = 2748.69$ m. The curves are derived from the 1D shear-wave velocity profiles predicted by the proposed PG-VAE (red solid lines), the unconstrained baseline dual encoder (blue solid lines), the shot-gather-only network (cyan dotted lines), the dispersion-image-only network (orange dotted lines), and the conventional 1D inversion (black dashed lines). The single-modality networks exhibit systematic velocity biases, and the unconstrained baseline network underestimates low-frequency phase velocities. The conventional inversion frequently aligns with the upper boundary of the high-energy band. The PG-VAE properly tracks the central high-energy trajectory of the fundamental mode.}
    \label{fig:real-picks-qc}
\end{figure}

To verify the physical consistency of the inverted velocity profiles, we forward-modeled their corresponding fundamental-mode dispersion curves and overlaid them onto the observed field spectral dispersion images (Figure \ref{fig:real-picks-qc}). 
The conventional 1D inversion (black dashed lines) captures the broad macroscopic trend but frequently deviates from the spectral maxima, systematically following the upper phase-velocity bound of the high-energy band rather than its center. This discrepancy underscores the subjective variability and inherent bias introduced by the manual dispersion-curve picking process, a key challenge in traditional inversions.
The unconstrained neural networks demonstrate significant weakness to domain shift. 
The single-modality ablation models exhibit systematic velocity bias on the complex field data. The shot-gather-only model (cyan dotted lines) systematically overestimates phase velocities across the entire analyzed frequency band, whereas the dispersion-image-only model (orange dotted lines) underestimates them, falling well below the fundamental mode's energy maxima. Even with multimodal inputs, the baseline dual encoder without physics-guided regularization (blue solid lines) still underestimates phase velocities at lower frequencies, resulting in physically inconsistent, low-velocity structural artifacts. 
In contrast, the dispersion curves derived from the PG-VAE (red solid lines) more consistently track the high-energy trajectory of the fundamental mode across all sampled locations. These results support that integrating forward-consistent physical constraints help data-driven architectures to move from synthetic training regimes to complex, real-world field applications.

\subsubsection{Geological Interpretation}

To translate the inverted $V_s$ profile into a practical geotechnical model, we apply standard near-surface seismic classification thresholds based on the National Earthquake Hazards Reduction Program (NEHRP) site classes \citep{NEHRP2020}. The velocity field is classified into three primary geological units. Velocities below 360 m/s are interpreted as soils or loose sand. The underlying intermediate layer, with velocities between 360 and 760 m/s, corresponds to very dense sand and weathered or soft rock. Finally, deeper regions with velocities exceeding 760 m/s denote competent rock \citep{NEHRP2020}.

\begin{figure}[htbp]
\centering
\includegraphics[width=\linewidth]{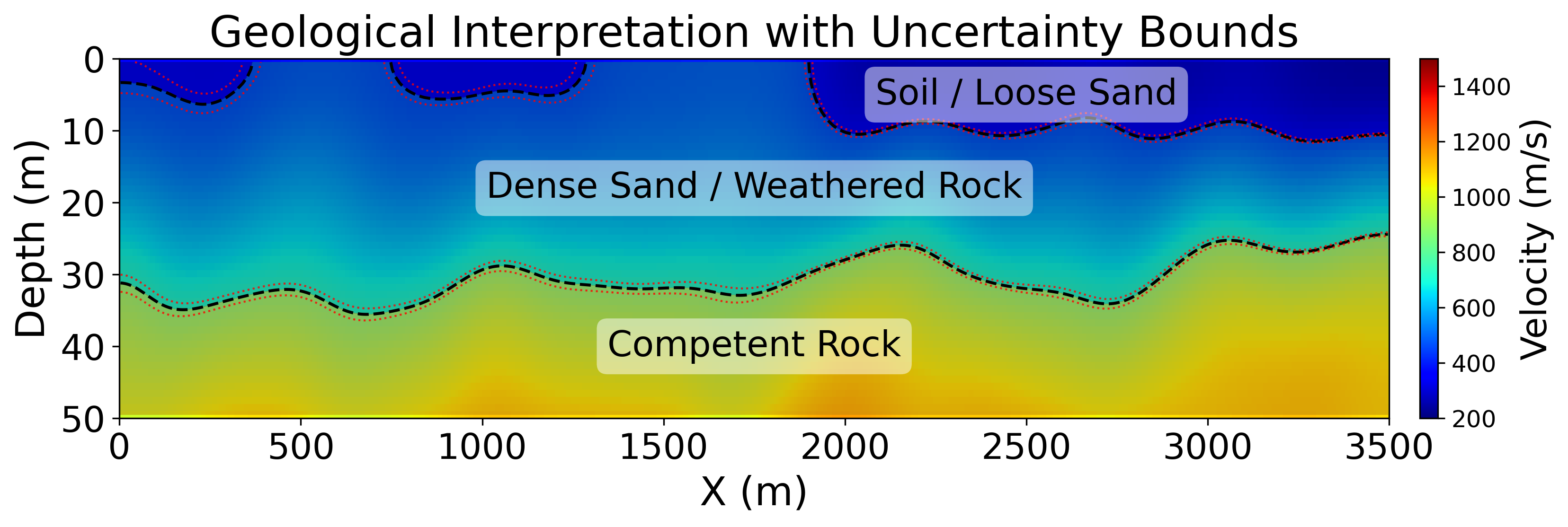}
\caption{Geological interpretation derived from the PG-VAE ensemble. The background color scale represents the continuous mean shear-wave velocity model ($V_s$ in m/s), transitioning from low velocities (cool colors) to high velocities (warm colors). The three geotechnical zones (soil / loose sand, dense sand / weathered rock, and competent rock) are categorized according to NEHRP classifications.
The central dashed black lines indicate the mean structural interfaces at 360 m/s and 760 m/s. The red dotted ribbons flanking these boundaries represent the approximate 99\% ($\pm 2.58\sigma$) confidence intervals, quantifying the network's increasing spatial uncertainty of the structural interfaces at greater depths.}
\label{fig:geo-interp2}
\end{figure}

Figure \ref{fig:geo-interp2} illustrates this interpretation, showing these classification zones on the mean velocity model. 
The color scale visually guides this transition, where cool blue hues denote the low-velocity shallow soil layer, transitioning through greens and yellows into the warmer orange and red hues of the competent rock at depth.
The dashed black contours mark the calculated mean spatial depth of the structural interfaces at the 360 m/s and 760 m/s thresholds.
To quantify the reliability of this interpretation, we leverage the PG-VAE's probabilistic nature to generate confidence intervals around these boundaries. The red dotted bands bracketing the mean interfaces represent the $\pm2.58\sigma$ interval, corresponding to approximately a 99\% confidence level, derived directly from the standard deviation. The ribbon widens along the highly undulating boundary between the weathered and competent rock, reflecting the diminishing sensitivity of surface waves at greater depths. This shows that the network successfully maps deeper, irregular structural geometries while providing critical uncertainty quantification for the interpreted layer boundaries.

\subsection{Robustness to random noise}

To quantitatively evaluate the robustness of the proposed method to noisy
data, we conducted a controlled synthetic experiment in which additive
random noise was introduced into the shot gathers. The signal-to-noise
ratio (SNR) was defined as

\begin{equation}
\mathrm{SNR}
=
10\log_{10}
\left(
\frac{\sum_{t,r} d(t,r)^2}
     {\sum_{t,r} n(t,r)^2}
\right),
\label{eq:snr}
\end{equation}

where $d(t,r)$ denotes the clean seismic data and $n(t,r)$ denotes the added
noise. Because the proposed network uses both the raw shot gather and its
spectral dispersion images as inputs, the dispersion image was
recomputed from each noise-contaminated shot gather rather than perturbing the
two inputs independently.

\begin{figure}
    \centering
    \includegraphics[width=\linewidth]{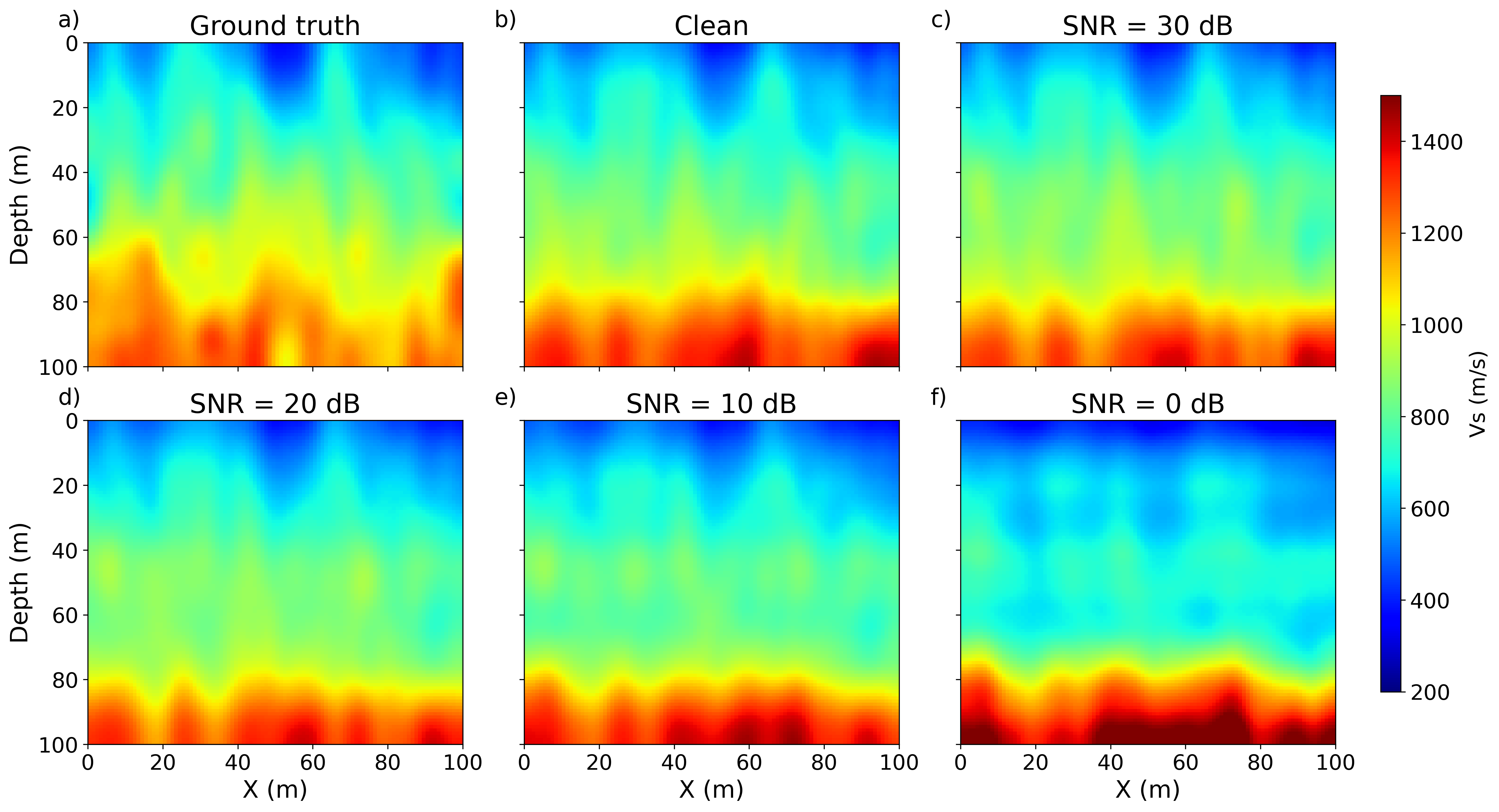}
    \caption{Results from noise-robustness evaluation of the fine-tuned PG-VAE. (a) Ground-truth $V_s$ model and mean PG-VAE
    predictions obtained using (b) clean data and shot gathers contaminated with
    additive random noise at SNR levels of (c) 30~dB, (d) 20~dB, (e) 10~dB,
    and (f) 0~dB. Noise was introduced in the shot-gather domain, and the
    corresponding dispersion image was recomputed from each noisy gather.
    During fine-tuning, each training sample had a 30\% probability
    of being contaminated with noise at an SNR randomly sampled between 10 and
    40~dB. The principal velocity structures remain comparatively stable down to
    10~dB, whereas substantial degradation is observed at 0~dB.}
    \label{fig:noise-pgvae}
\end{figure}

Initial tests showed that the PG-VAE trained exclusively on clean synthetic
data was sensitive to noise-induced distribution shift. We therefore introduced
stochastic noise augmentation during a short fine-tuning stage. Starting from
the weights obtained from clean-data training, each training sample had a
30\% probability of being contaminated with random noise. For noisy
samples, the SNR was randomly drawn between 10 and 40~dB. The PG-VAE was
fine-tuned using the same optimization strategy as the original network and
converged after 8 additional epochs based on the early-stopping criterion.
Figure~\ref{fig:noise-pgvae} shows the resulting PG-VAE predictions for
clean input data and for SNR levels of 30, 20, 10, and 0~dB. The principal subsurface interfaces remain visually stable for SNR values down to approximately 10~dB. However, the accuracy progressively decreases as the noise level increases. Table~\ref{tab:noise-pgvae} summarizes the quantitative metrics before applying the Gaussian smoothing to the models. The MAE increases
from 126.06~m/s for the clean data to 136.40, 143.96, and 163.05~m/s at
30, 20, and 10~dB, respectively. The corresponding $R^2$ score decreases from
0.6467 for the clean case to 0.4227 at 10~dB. At 0~dB, the
structural detail is significantly reduced, with the MAE increasing to 219.57~m/s and
the $R^2$ score decreasing to 0.0630.

\begin{table}[htbp]
\centering
\begin{tabular}{|c|c|c|c|c|c|}
\hline
{SNR} & {MAE (m/s) $\downarrow$} & {RMSE (m/s) $\downarrow$} & {MAPE (\%) $\downarrow$} & {$R^2$ $\uparrow$} & {SSIM $\uparrow$} \\
\hline
{Clean} & {126.06} & {181.53} & {14.78} & {0.6467} & {0.6565} \\
{30 dB} & {136.40} & {197.64} & {15.70} & {0.5812} & {0.5694} \\
{20 dB} & {143.96} & {209.10} & {16.56} & {0.5312} & {0.5350} \\
{10 dB} & {163.05} & {232.05} & {18.88} & {0.4227} & {0.4293} \\
{0 dB}  & {219.57} & {295.62} & {25.19} & {0.0630} & {0.2486} \\
\hline
\end{tabular}
\caption{Quantitative evaluation of the PG-VAE on the noisy
synthetic test model. The PG-VAE was fine-tuned using stochastic noise augmentation with
training SNR values sampled between 10 and 40~dB.}
\label{tab:noise-pgvae}
\end{table}

As an additional control, the same stochastic noise-augmentation strategy was
applied to the baseline VAE without the physics-guided loss. The baseline
network exhibited a comparable improvement in tolerance to additive noise,
although the PG-VAE retained lower MAE and MAPE values from the clean case
through 10~dB. For comparison, the baseline MAE increased from 139.12~m/s for clean data
to 173.27~m/s at 10~dB and 225.62~m/s at 0~dB.
This comparison indicates that the improved tolerance to
input-domain noise arises primarily from explicit noise augmentation. 
The role of the physics constraint is
instead to regularize the predicted velocity models toward solutions that remain
consistent with the modeled surface-wave dispersion physics. These results distinguish robustness to input-data perturbations from
physics-based regularization of the inverse solution. 
To verify that fine-tuning did not substantially alter the field-data
interpretation, we also applied the fine-tuned PG-VAE to the Devine dataset.
The resulting $V_s$ model remained qualitatively consistent with that obtained
using the original clean-trained network, preserving the principal shallow and
deep structural features.

\subsection{Limitations and future work}

Although this study demonstrates the effectiveness of the PG-VAE framework for near-surface characterization, several limitations need to be acknowledged. The physics constraint imposed by the surrogate model relies on assumptions about a 1D vertical profile. While this 1D approximation is a standard and computationally efficient approach for surface-wave dispersion analysis, it inherently neglects complex 3D scattering, multipath effects, and out-of-plane propagation that are often present in field data. Although our pseudo-2D inference approach yields spatially coherent results, it cannot fully account for strong lateral heterogeneities. The current inversion framework solves exclusively for S-wave velocity ($V_s$), treating P-wave velocity ($V_p$) and bulk density ($\rho$) as fixed empirical functions. We assume a constant $V_p/V_s$ ratio of 2 and density using Gardner's relation ($0.31V_p^{0.25}$). However, this approximation may introduce structural artifacts in complex near-surface environments, such as unconsolidated sediments or fluid-saturated zones, where these physical parameters are frequently decoupled. The network's generalization capability is inherently tied to the synthetic training distribution. While we introduced stochastic perturbations to mimic geological variability, real seismic data contain complex source-wavelet variations, irregular topography, and coherent cultural noise, all of which are difficult to simulate accurately. This simulation-to-reality gap can substantially degrade performance when strong input-domain noise or acquisition effects are absent from the training
distribution. The controlled experiments presented here show that explicit fine-tuning improves tolerance to additive noise. Future research should focus on integrating fully elastic 2D or 3D forward operators into the surrogate loop to directly invert for lateral heterogeneities. Additionally, expanding the latent space to incorporate multi-parameter inversion for $V_p$ and density, and explicitly handling higher-order Rayleigh wave modes, will be essential steps to further reduce model uncertainties and improve the fidelity of subsurface characterization.

Because our physics guidance relies on a data-driven surrogate model rather than a true analytical or numerical forward operator embedded directly within the training loop, the framework is subject to a compounded generalization risk. The surrogate approximates wave physics strictly within the bounds of its synthetic training distribution. Consequently, any out-of-distribution errors, biases, or inaccuracies learned by the surrogate will inherently propagate into the main VAE during training. The forward-consistency constraint serves as a learned physical regularizer, but it cannot enforce strict physical validity if the predicted model lies outside the surrogate's generalization bounds. Replacing the neural surrogate with a fully differentiable numerical PDE solver remains an important objective for future research to provide strict physical guarantees.

\section{Conclusion}

We introduced a physics-guided variational autoencoder (PG-VAE) for the automated, probabilistic inversion of near-surface shear-wave velocity ($V_s$) profiles from raw seismic shot gathers and spectral dispersion images. By incorporating a differentiable physics surrogate into the training loop, we enforce forward-consistency between the predicted velocity structures and the observed dispersive wave physics. Validation on synthetic datasets demonstrated that an unconstrained data-driven network is prone to producing structurally incorrect results, whereas PG-VAE reliably recovers complex, undulating subsurface structures with high fidelity. Controlled synthetic experiments quantify the sensitivity of the network to additive noise, while application to the Devine dataset evaluates its transfer to field observations. The physics constraint effectively served as a regularizer, successfully extracting coherent physical features, whereas the baseline network produced results less consistent with the dispersion behavior. The probabilistic nature of the VAE yields an ensemble of solutions, providing spatially localized uncertainty maps. This uncertainty quantification provides explicit confidence intervals in deeper or highly heterogeneous layers, where surface-wave sensitivity degrades. The physics-guided framework reduces subjective bias and computational bottlenecks associated with manual dispersion-curve picking, offering a scalable, reliable, and interpretable solution for large-scale near-surface seismic characterization.

\section{ACKNOWLEDGMENTS}

We acknowledge the sponsors of the Texas Consortium for Computational Seismology (TCCS) and GeoH2 industrial consortia at the Bureau of Economic Geology for the financial support. 
We thank Alex Kaminsky for the software and for discussions on conventional surface-wave inversion strategies. We also acknowledge Keurfon Luu for developing and maintaining the open-source Python packages disba (https://github.com/keurfonluu/disba) and evodcinv (https://github.com/keurfonluu/evodcinv), which were used for dispersion modeling and inversion in this study.

\bibliographystyle{unsrt}  
\bibliography{references}

\end{document}